\documentclass[10pt,prb,aps,twocolumn,superscriptaddress,floatfix,longbibliography]{revtex4-2}

\usepackage{amsmath,amssymb,amsfonts,bbm,color,graphicx,tikz}

\usepackage{ragged2e}
\usepackage[normalem]{ulem}
\usepackage[utf8]{inputenc}
\usepackage[T1]{fontenc}
\usepackage[version=4]{mhchem} 

\usepackage{wasysym}

\usepackage[unicode=true,colorlinks=true]{hyperref}
\hypersetup{linkcolor=blue,citecolor=blue,urlcolor=blue}

\usepackage{bm}
\usepackage{dsfont}

\usepackage{braket}

\newcommand{\iu}{\mathrm{i}} 
\newcommand{\eu}{\mathrm{e}} 

\newcommand{\bvec}[1]{\bm{#1}}

\newcommand{\Ztwo}{\mathbb{Z}_2}

\newcommand{\SUthree}{\mathrm{SU(3)}}

\newcommand{\SOtwo}{\mathrm{SO(2)}}
\newcommand{\SOthree}{\mathrm{SO(3)}}

\newcommand{\mycomment}[1]{}

\newcommand{\Qzeigenone}{%
    \tikz[baseline=0.1cm]{%
        \draw[solid, thick] (0,0) -- (0.4,0.4);
        \draw[solid, thick, rotate around={-45:(0.2,0.2)}] (0.2,0.2) ellipse (0.15 and 0.05);
    }%
}

\newcommand{\Qzeigentwo}{%
    \tikz[baseline=0.1cm]{%
        \draw[solid, thick] (0,0.40) -- (0.40,0);
        \draw[solid, thick, rotate around={45:(0.20,0.2)}] (0.2,0.2) ellipse (0.15 and 0.05);
    }%
}

\newcommand{\Qzeigenthree}{%
    \tikz[baseline=0.1cm]{%
        \draw[solid, thick, ] (0.2,0.2) circle (0.15);
        \filldraw[color = black] (0.2,0.2) circle (0.02)
    }%
}

\newcommand{\uvec}[1]{{\underline{#1}}}

\DeclareMathOperator{\diag}{diag}

\begin{document}

\title{Exchange-frustrated quadrupoles on the honeycomb lattice: Flavor-wave spectra, classical degeneracies and parton constructions}

\author{Partha Sarker}
\affiliation{Institute for Theoretical Physics, University of Cologne, 50937 Cologne, Germany}
\author{Han Ma}
\affiliation{CPHT, CNRS, Ecole Polytechnique, Institut Polytechnique de Paris, Palaiseau 91128, France}
\affiliation{Department of Physics and Astronomy, Stony Brook University, Stony Brook, New York 11974, USA}
\affiliation{Perimeter Institute for Theoretical Physics, Waterloo ON N2L 2Y5, Canada}
\author{Urban F.~P.~Seifert}
\affiliation{Institute for Theoretical Physics, University of Cologne, 50937 Cologne, Germany}

\begin{abstract}
We study the quadrupolar Kitaev model, an $S=1$ honeycomb-lattice model with frustrated bond-dependent quadrupolar interactions.
Using complementary methods and expanding around controlled limits, we uncover several intertwined structures. First, a semiclassical variational analysis based on $\mathrm{SU}(3)$ flavor theory reveals an extensively degenerate manifold of classical mean-field ground states, suggesting that quantum fluctuations may stabilize a quantum-disordered phase.
Second, in the bond-anisotropic limit, perturbation theory is used to derive effective low-energy Hamiltonians, which crucially depend on the presence (or absence) of a residual symmetry $\mathcal{M}$ of combined lattice reflection and discrete spin rotation.
A Majorana parton construction uncovers an exact $\mathbb Z_2$ gauge structure and motivates possible confined and deconfined phases driven by gauge-charge condensation, consistent with the effective theories obtained in anisotropic limit.
Further, within the same parton formalism, different Majorana mean-field ansätze produce both gapless and gapped candidate quantum-disordered states, distinguished by linear versus projective implementations of $\mathcal M$.  Our results highlight frustrated quadrupolar interactions as a route to quantum-disordered phases, relevant to $S \geq 1$ Kitaev materials and Rydberg-array quantum simulators.
\end{abstract}

\date{\today}

\maketitle
\section{Introduction}

Frustrated magnets are a fruitful venue for realizing and exploring \emph{quantum} phases of matter: frustration suppresses classical ordering tendencies, and opens a route for quantum fluctuations to stabilize non-trivial ground states, in particular Quantum Spin Liquids (QSLs). Here, spins form a highly entangled state that supports fractionalized excitations and is described by an emergent gauge theory in a deconfined phase \cite{savaryQuantumSpinLiquids2016,knolleFieldGuideQSL2019}.
A paradigmatic example is the Kitaev model \cite{kitaevAnyonsExactlySolved2006,hermannsPhysicsKitaevModel2018,takagiConceptRealization2019} of spin-1/2 local moments on the honeycomb lattice, owing to its exact solvability \cite{kitaevAnyonsExactlySolved2006} and to its approximate realization in spin-orbit coupled Mott insulators with additional non-Kitaev interactions in real materials \cite{jackeliMottInsulators2009,yokoiHalfInteger2021,sarkisIntermediateField2026}.

Concurrently, recent years have seen renewed interest in systems hosting interacting multipolar degrees of freedom \cite{nakatsujiSpinDisorder2005,laeuchliQuadrupolar2006,szaszPhaseDiagram2022,pohleSpinNematicsMeet2023}. 
They arise naturally in systems with local moments of spin $S \geq 1$ and can lead to time-reversal-even states that nonetheless break spin-rotation symmetry: while no spin polarization is present, $\braket{\vec{S}} = 0$, \emph{spin fluctuations} as characterized by $\braket{S^\alpha S^\beta}$ can break the $\SOthree$ symmetry of spin rotations down to O(2) (for uniaxial spin nematics).
Ordering of such degrees of freedom thus allows for spin-nematic phases with ``hidden'' multipolar order parameters \cite{pencSpinNematicPhases2011,jiangWhereIsThe2023}.
Even in a system of $S\geq 1$ exhibits dipolar order, its dynamical response may carry fingerprints of quadrupolar excitations that are not described by conventional spin wave methods \cite{baiHybridizedQuadrupolarExcitations2021}.

Given that QSLs are typically pursued in systems where frustration suppresses \emph{dipolar} ordering tendencies of $S=1/2$ moments \cite{savaryQuantumSpinLiquids2016,knolleFieldGuideQSL2019}, it is an interesting question to ask to what extent frustrated interactions can suppress multipolar order \cite{seifertPhaseDiagramsExcitations2022,chungGeometricallyFrustratedQuadrupoles2025} and instead stabilize quantum-liquid states of multipolar degrees of freedom, and how these states can be characterized.
Frustrated interactions between multipolar moments have also attracted interest from an experimental perspective: on the one hand, exchange-coupled local moments of non-Kramers ions naturally exhibit a mixed multipolar character, with rare-earth pyrochlores such as Tb$_2$Ti$_2$O$_7$ being a prominent example \cite{chungGeometricallyFrustratedQuadrupoles2025,rauFrustratedQuantum2019,guittenyAnisotropic2013}, and recent works have pointed out that d$^2$ spin-orbit Mott insulators may realize \emph{bond-dependent} multipolar interactions \cite{rayyanFieldInduced2023,khaliullinExchangeInteractions2021}.
On the other hand, there has been significant progress in simulating models with larger local Hilbert spaces, which can be mapped onto $S \geq 1/2$ degrees of freedom, using Rydberg array quantum simulators \cite{liuSuperSoliditySimplex2024,verresenPredictionToricCode2021b,verresenUnifyingKitaevMagnets2022,wangRenormalizedClassical2025}. 

Motivated by above developments, in this work, we consider a quadrupolar version of Kitaev's honeycomb model, wherein local $S=1$ moments interact via bond-dependent quadrupolar interactions.
This model was derived in Ref.~\onlinecite{verresenUnifyingKitaevMagnets2022} as an effective model for Ruby lattice Rydberg atom arrays \cite{semeghiniProbingTopologicalSpin2021}, and numerical simulations point towards a gapped $\Ztwo$ topologically ordered ground state.
Given a by-now deep understanding of Kitaev models with dipolar bond-dependent interactions (both for $S=1/2$ and general $S$) \cite{hermannsPhysicsKitaevModel2018,takagiConceptRealization2019,khaitCharacterizing2016,maZ2HigherSpin2023}, a quadrupolar version of Kitaev's honeycomb model thus appears as a valuable drosophila for studying novel phases in systems of frustrated multipoles.
To this end, we deform the Hamiltonian by a Zeeman field or single-ion anisotropy which act as polarizing fields for dipolar (spins) and quadrupolar degrees of freedom, respectively.
Upon increasing the strength of Kitaev interactions, the trivial polarized states eventually become unstable, leading to a possible spin-liquid regime with an extensively degenerate mean-field ground-state manifold.
Along another direction in parameter space, we start from the anisotropic limit with $J_z \gg J_{x,y}$, where we can use perturbation theory to derive effective Hamiltonians. We highlight the crucial role of a combined reflection and spin-rotation symmetry $\mathcal{M}_z$ for $J_x =J_y \ll J_z$, where the leading anisotropic effective Hamiltonian yields an (sub-)extensively degenerate manifold rather than a simple trivial product state.
By introducing a Majorana parton construction, we identify an \emph{exact} $\Ztwo$ gauge structure and discuss characteristic string operators as diagnostics for deconfined and confined phases, with and without broken $\mathcal{M}_z$.
We complement this analysis by self-consistent Majorana mean-field theory calculations, where we find that the projective implementation of the $\mathcal{M}_z$-symmetry enables a gapped mean-field ground state.

The remainder of the paper is organized as follows. 
In Sec.~\ref{sec:modelsymm}, we introduce the model and discuss relevant symmetries.
We investigate semiclassical ground states and their degeneracy in Sec.~\ref{sec:semiclassics}.
An effective (quantum) Hamiltonian in the limit of highly anisotropic couplings is derived in Sec.~\ref{sec:anisotropy-limit}.
We analyze the model's exact gauge structure using a parton construction in Sec.~\ref{sec:exact-partons}, and present results from parton mean-field calculations in Sec.~\ref{sec:mmft}.
A discussion (Sec.~\ref{sec:conclusion}) closes the paper.

\section{The quadrupolar Kitaev model} \label{sec:modelsymm}

\subsection{Hamiltonian}

\begin{figure}[tbp]
\centering
\includegraphics[width=\columnwidth]{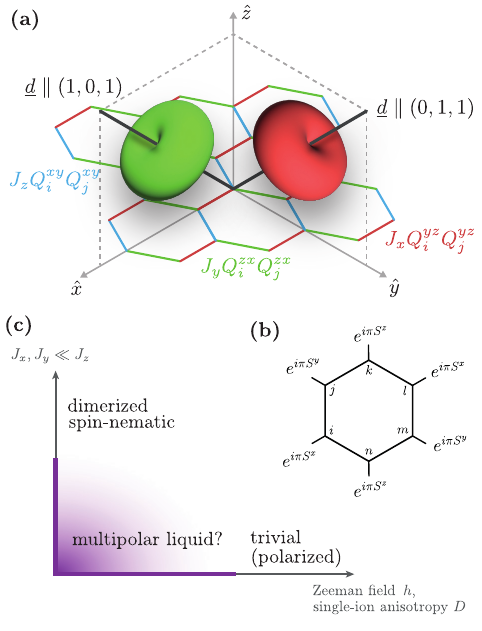}
\caption{(a) Illustration of frustrated bond-dependent interactions on the honeycomb lattice as described by the Hamiltonian in Eq.~\eqref{eq:Hamil}. For example, the antiferroquadrupolar interaction on the $y$-bond may favor a spin-nematic state with a director along the $(1,0,1)$ axis, competing with an antiferroquadrupolar interaction on the $x$-bond which can be maximized by a director along the $(0,1,1)$ direction. The red and green tori indicate the probability distribution for a measurement to observe the spin in the respective direction,~i.e.~ $|\braket{S(\theta,\varphi)|\uvec{d}}|^2$ with $\ket{S(\theta,\varphi)}$ being a spin-coherent state. (b)
Illustration of hexagonal plaquette operator $W_{\mathcal{L}}=\prod_{i \in \partial \mathcal{L} = \varhexagon} \exp(\iu \pi S_i^{\bar\alpha})$, which can be extended to any closed loop in the honeycomb lattice. (c) Schematic phase diagram, where a Zeeman field, single-ion anisotropy, as well as bond-anisotropic interactions, are deformations that establish solvable limits which we use as starting points for controlled insights. 
}
\label{fig:lattice,AFQ order,plaquette}
\end{figure}

We consider spin-$1$ degrees of freedom on the honeycomb lattice as shown in Fig.~\ref{fig:lattice,AFQ order,plaquette}, which interact via bond-dependent quadrupolar interactions of strengths $J_x, J_y, J_z > 0$ on $x$-,$y$- and $z$-bonds, as described by the Hamiltonian \cite{verresenUnifyingKitaevMagnets2022}
\begin{equation} \label{eq:Hamil}
	H_Q = J_x \sum_{\langle ij \rangle_x} Q_{i}^{yz} Q_{j}^{yz} + J_y \sum_{\langle ij \rangle_y} Q_{i}^{zx} Q_{j}^{zx} + J_z \sum_{\langle ij \rangle_z} Q_{i}^{xy} Q_{j}^{xy}.
\end{equation}
Here, we have introduced the quadrupolar operators $Q^{\alpha\beta}$ which transform as rank-2 traceless symmetric tensors under spin rotations, and can be expressed in terms of the spin operators $S^\alpha$ as
\begin{equation} \label{eq:Qopdefinition}
    Q^{\alpha\beta} = S^{\alpha}S^{\beta}+S^{\beta}S^{\alpha}-\frac{4}{3}\delta^{\alpha\beta}.
\end{equation}
Note that there exist five \emph{linearly independent} quadrupolar operators, with a convenient basis set given by $\{ Q^{xy}, Q^{yz}, Q^{zx}, Q^{x^2 - y^2},Q^{3z^2 - r^2} \}$ \footnote{
$Q^{x^2-y^2}=\frac{1}{2}(Q^{xx}-Q^{yy})$ and $Q^{3z^2 - r^2}=-\frac{1}{2\sqrt{3}}(Q^{xx}+Q^{yy}-2Q^{zz})$} which can be thought as a five component-vector $\bvec{Q}$.

In the remainder of this paper, for explicit calculations it is useful to adopt a time-reversal invariant basis $\ket{x},\ket{y},\ket{z}$ of the local Hilbert space, introduced in Appendix~\ref{app:d-vec-formalism}. In this basis, any local state can be written as $\ket{d} = \sum_{\alpha=x,y,z} d^\alpha \ket{\alpha}$, where the normalized vector $\uvec{d} \in \mathbb{C}^3$ is defined up to an overall phase. Real $\uvec{d}$ corresponds to a purely spin-nematic state, while complex $\uvec{d}$ can in general signify a finite dipolar moment.

Using this basis, the frustrated nature of $H_Q$ is readily demonstrated.
For example, the interaction on the $x$-bonds $\sim Q^{yz}_i Q^{yz}_{i+\hat{x}}$ favours anti-aligned directors $d_i = (0,1,1)^\top/\sqrt{2} $ and $d_{i+\hat{x}} =  (0,-1,1)^\top / \sqrt{2}$ (or vice versa) perpendicular to the $x$-axis in spin space, while on the $y$-bond the interaction $\sim Q^{zx}_i Q^{zx}_{i+\hat{y}}$ is minimized by $d_{i} = (1,0,1)^\top / \sqrt{2}$ and $d_{i+\hat{y}} = (-1,0,1)^\top / \sqrt{2}$ (or vice versa), as illustrated in Fig.~\ref{fig:lattice,AFQ order,plaquette}(a).
Hence, the local quadrupolar degrees of freedom (as specified by the directors) are exchange-frustrated, as common in systems with compass-type interactions \cite{nussinovCompassModels2015}. 

\subsection{Symmetries of the model} \label{subsec:symm}

We list pertinent symmetries of the quadrupolar Hamiltonian, which will guide the analysis of different deformations of $H_Q$ and the construction of symmetric mean-field ansätze.

\emph{First,} as $H_Q$ only contains quadrupolar operators, the  model is manifestly time reversal symmetric 
\begin{equation}
    Q^{\alpha \beta} \xrightarrow{\mathcal{T}} Q^{\alpha \beta},
\end{equation}
where $\mathcal{T}: \vec{S} \to -\vec{S}$ acts as complex conjugation in the basis for $S=1$ as chosen above.

\emph{Second,} there exists a global $\mathbb{Z}_2 \times \mathbb{Z}_2$ (dihedral) symmetry of spin rotations by $\pi$ along thee cartesian axis $\alpha = x,y,z$ (note only two out of the three spin rotations are independent). These act as $S^\beta \to \eu^{\iu \pi S^\alpha} S^\beta e^{-\iu \pi S^{\alpha}} = - S^\beta (+S^\beta)$ for $\beta \neq \alpha$ ($\beta = \alpha$), and the quadrupolar operators transform as \begin{equation}
\label{eq:spin_rotation_Qops_transform}
    \eu^{\iu \pi S^\alpha } Q^{\beta \gamma} \eu^{-\iu \pi S^\alpha } \to - Q^{\beta \gamma} (+Q^{\beta \gamma})
\end{equation} for $\alpha = \beta $ or $\alpha = \gamma$ ($\alpha \neq \beta$ and $\alpha \neq \gamma$).

\emph{Third}, as in dipolar Kitaev models (for arbitrary spin $S$) \cite{baskaranSpinSKitaevModel2008}, the model possesses an extensive number of conserved operators
\begin{equation}
	W_\mathcal{L} = \prod_{i \in \mathcal{L}} \eu^{\iu \pi S^{\alpha(i)}_i}
\end{equation}
given by site-dependent spin rotations by $\pi$ around the axis $\alpha$ corresponding to the bond type emanating site $i$ which is \emph{not} part of the closed loop $\mathcal{L}$ of the honeycomb lattice, thereby giving rise to a $\Ztwo$ one-form symmetry \cite{maZ2HigherSpin2023,liuSymmetriesAnomalies2024}.
A particularly important class are the \emph{plaquette operators} $W_{\partial p}$ associated with the closed loops around the hexagonal plaquette $p$, as shown in Fig.~\ref{fig:lattice,AFQ order,plaquette}.
We further note the idempotent nature of $(W_\mathcal{L})^2=1$, implying that they have eigenvalues $\pm 1$.

\emph{Finally,} the model retains the translational symmetry of the honeycomb lattice. Due to the bond-dependent character of interactions, point group symmetry operations become locked to suitable rotations in spin space:
\begin{enumerate}
	\item For isotropic couplings, $J_x = J_y=J_z$, there is $C_6$ lattice rotation symmetry combined with a spin rotation about the [111] axis, which maps $Q^{yz} \to Q^{zx} \to Q^{xy}$.
	\item If at least two of the three coupling constants are equal, e.g. $J_x = J_y \neq J_z$, the system retains a mirror symmetry about an axis parallel to the $z$-bonds, combined with $\pi/2$ rotation in spin space about the same axis (here: about the $z$-axis, which maps $Q^{xy} \to - Q^{xy}$, $Q^{yz} \to - Q^{zx}$ and $Q^{zx} \to Q^{yz}$).
\end{enumerate}

\section{Solvable limits\label{sec:solvable_limits}}

To gain insight into the ground-state structure of the quadrupolar Hamiltonian $H_Q$, we adopt the general strategy of approaching it from controlled limits in which the many-body ground state is more transparent. One such route is to consider a deformed Hamiltonian $H_Q + H'$, where the deformation $H'$ is chosen such that the strong-deformation limit is solvable and admits a reference ground state that can be characterized in a reliable way.
Upon reducing the deformation, the quadrupolar interactions in $H_Q$ progressively reshape the spectrum. A closing of the quasiparticle gap at a critical deformation strength then signals an \emph{instability} of the reference state.
As a caveat, we note that this instability may in principle be pre-empted by a first-order phase transition,
or the reference state may remain a (but not necessarily
the only) ground state all the way to vanishing deformation strength.

Further insight is gained by examining the anisotropic limit of $H_Q$ itself, in which one of the quadrupolar couplings dominates and the ground state reduces to a simple product state.
As in the Kitaev honeycomb model, beyond the local physics, this limit can reveal the effective gauge structure and global topological properties of a phase connected to the isotropic regime.

\begin{enumerate}
	\item Field-polarized limit: We consider a magnetic field $h>0$ along the $[1,1,1]$ axis,
	\begin{equation}\label{eq:deformation h}
		H'_h=-h\sum_i \left(S_i^x+S_i^y+S_i^z\right).
	\end{equation}
	In the limit of $h/J^\alpha\to\infty$ for all $\alpha$, the ground state is the product state in which each spin is fully polarized along $[111]$ direction. In the $d$-vector basis (see App.~\ref{app:d-vec-formalism}),
	\begin{equation}
		\ket{\psi_g}=\prod_i \ket{d}_i,
	\end{equation}
	with $\uvec d = (-1+\iu\sqrt{3},\, -1-\iu\sqrt{3},\, 2)^\top/(2\sqrt{3})$.
	In this state, the spins are polarized as $\langle \vec S_i\rangle = \frac{1}{\sqrt{3}}(1,1,1)^\top$,
	and the quadrupolar expectation values are $\langle \bvec Q_i\rangle = \frac{1}{3}(1,1,1,0,0)^\top$.
	\item Strong single-ion anisotropy limit: We further deform $H_Q$ by a single-ion anisotropy along the $[111]$ axis,
	\begin{align}\label{eq:deformation D}
    	H'_D
    	&= D \sum_{i} \left(S_i^x + S_i^y + S_i^z\right)^2 \nonumber\\
    	&\equiv D\sum_i \left(Q_i^{xy}+Q_i^{yz}+Q_i^{zx}+2\right).
 	\end{align}
    For $D>0$, the deformation favors the product state given by $\ket{\psi_g}=\prod_i \ket{d}_i$,
 	with $\uvec d=(1,1,1)^\top/\sqrt{3}$.
 	This is a purely quadrupolar state, with $\langle \vec S_i\rangle = 0$ and $\langle \bvec Q_i\rangle = -\frac{1}{3}(2,2,2,0,0)^\top$.
    We emphasize that, whereas for classical dipolar spins $D>0$ corresponds to an easy-plane anisotropy with a $U(1)$-degenerate manifold of spin configurations, the quantum-mechanical ground state of this deformation is unique.
    The single-ion anisotropy can therefore be viewed as a polarizing field for a spin-nematic state for $D>0$.
    Both field-deformation and single-ion anisotropy will be studied in Sec.~\ref{sec:semiclassics}.

	\item Strong bond-anisotropic limit: We consider $H_Q$ with unequal couplings $J^\alpha$, which breaks the the cubic symmetry of the isotropic point $J^x=J^y=J^z$. In the limit of $J^\gamma \gg J^{\alpha \neq \gamma}$, the model reduces at leading order to decoupled $\gamma$ bonds, so the many-body ground state is a product of strong-bond ground states. For example, when $|J^z| \gg |J^{x,y}|$, each strong bond (connecting sites $i$ and $j$) satisfies $Q^{xy}|d_\pm \rangle = \pm |d_\pm \rangle$, given by $|d_\pm \rangle =(\mp 1,1,0)^\top/\sqrt{2}$. For $J^z<0$, the bond ground states are $|d_\pm\rangle_i |d_\pm\rangle_j$ whereas for $J^z>0$, they are $|d_\pm\rangle_i |d_\mp\rangle_j$. This is directly analogous to the well-known strong-anisotropy limit of the Kitaev honeycomb model, which provides the starting point for perturbative constructions of the low-energy effective theory. The detailed study of this limit is in Sec.~\ref{sec:anisotropy-limit}.
\end{enumerate}

\section{Semi-classical Flavor-Wave Analysis and instabilities of the deformed quadrupolar model} \label{sec:semiclassics}

\subsection{Instability of field-polarized and polarized-quadrupolar states} 

We systematically study the excitation spectra above the reference states introduced above, as well as their putative instabilities. In particular, in this section we consider the two onsite deformations introduced above, namely the $[111]$-field and the single-ion anisotropy. We use $\SUthree$ flavor-wave theory \cite{pencSpinNematicPhases2011,munizGeneralizedSpinwaveTheory2014}, which can be viewed as a generalized form of Holstein-Primakoff spin-wave theory. Whereas conventional spin-wave theory assumes a classically ordered dipolar background and elementary excitations with $\Delta S=1$, flavor-wave theory is designed to also describe fluctuations about quadrupolar, or spin-nematic, states.

To this end, we introduce three Schwinger bosons $b_\gamma$, with $\gamma\in\{x,y,z\}$, and represent the spin-1 operators as $ S^{\alpha} = -\iu \epsilon^{\alpha \beta \gamma}  b_{\beta}^{\dagger} b_{\gamma}$ along with the constraint  $\sum_{\alpha}b_{\alpha}^{\dagger}b_{\alpha} =M$ with the physical spin-1 Hilbert space recovered at $M =1$. For book-keeping purposes, it is convenient to leave $M$ as a free parameter \cite{pencSpinNematicPhases2011,munizGeneralizedSpinwaveTheory2014}. In this representation, the quadrupolar operators can be written as bilinear forms 
\begin{equation} Q_i^{\alpha\beta} = \left(\begin{array}{ccc}
 b_{i,x}^{\dagger} & b_{i,y}^{\dagger} & b_{i,z}^{\dagger}
\end{array} \right)\underline{\underline{Q}}^{\alpha\beta} \left(\begin{array}{ccc}
 b_{i,x} \\ b_{i,y} \\ b_{i,z}
\end{array} \right),
\end{equation}
where $\underline{\underline{Q}}^{\alpha\beta}$ are corresponding $3\times 3$ matrices given in Appendix~\ref{app:d-vec-formalism}.

At the mean-field level, boson condensation \(b_\gamma \to \langle b_\gamma\rangle = d_\gamma\) directly reproduces the \(d\)-vector description of the local state, since
\[
\langle S^\alpha\rangle
=
-\iu\,\epsilon^{\alpha\beta\gamma} d_\beta^* d_\gamma
=
\bra d S^\alpha \ket d.
\]
To study fluctuations about the condensate, one performs a generalized Holstein-Primakoff expansion. This is simplest in a canonical frame in which the condensate is aligned with $\uvec d_{\rm can.}=(0,0,1)^\top$
so that the \(z\)-flavor is macroscopically occupied and the remaining two flavors describe fluctuations.
In that frame, one eliminates the condensed boson according to
\begin{equation}
b_z',\, b_z^{\prime\dagger}\to \sqrt{M-b_x^{\prime\dagger}b_x'-b_y^{\prime\dagger}b_y'}.
\end{equation}
For a general reference state, one first performs a unitary transformation to such a canonical frame. Then the bosons in the laboratory and canonical frames are related by
$
\uvec b = \underline{\underline{U}} \uvec b'
$
and condensation in the primed basis,
$
\langle \uvec b' \rangle \propto (0,0,1)^\top
$,
reproduces the desired condensate in the original basis (see Appendix~\ref{app:lfwt} for details),
$
\langle \uvec b \rangle \propto \underline{\underline{U}} (0,0,1)^\top = \uvec d
$.
The generalized Holstein-Primakoff expansion is then carried out in the primed basis.

With this, one obtains a controlled large-\(M\) expansion of the deformed Hamiltonian,
\begin{equation}\label{eq:lfwt-expansion}
    H_Q+H' = M^2 H^{(0)} + M H^{(2)} + \mathcal{O}\!\left(M^{1/2}\right),
\end{equation}
where \(H^{(0)}\) is the mean-field energy of the reference product state \(\prod_i \ket{d}_i\) which is a number, and \(H^{(2)}\) is quadratic in the fluctuation bosons.
Diagonalizing \(H^{(2)}\) yields the linear flavor-wave spectrum. Higher-order terms, which are subleading in \(1/M\), describe interactions between flavor-wave excitations and are neglected within linear flavor-wave theory.

\begin{figure}[tbp]
    \centering
    \includegraphics[width=\columnwidth]{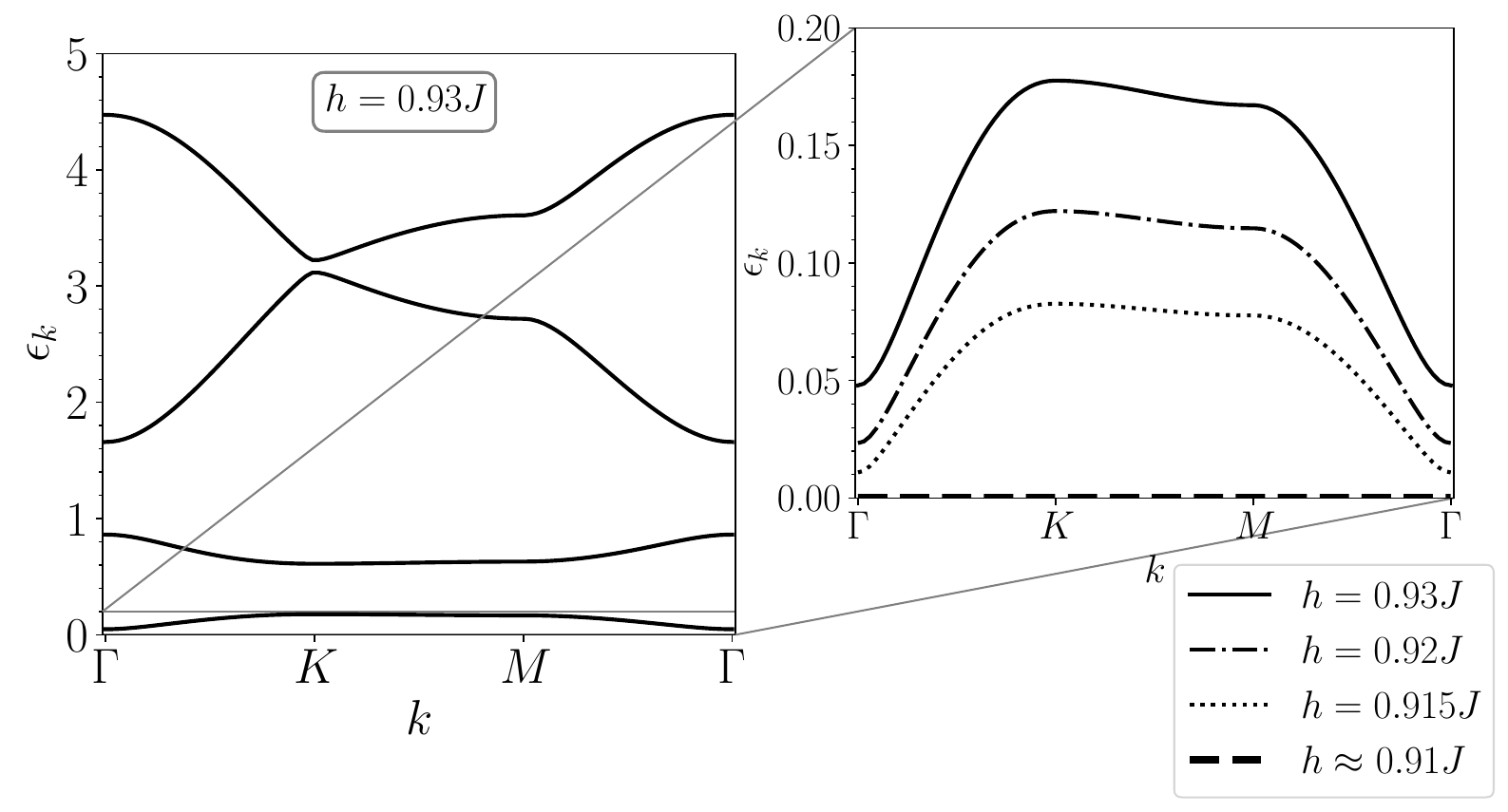}
    \caption{The band structure of flavor-wave excitations on top of the spin-polarized reference state evolves as the magnetic field lowered. At $h \approx 0.91 J$, the flavor wave excitations become gapless and obtain a macroscopic occupation at all momenta,~i.e.~$\langle b_k \rangle = \mathrm{constant}$. As a result, the polarized reference state for a Zeeman field along the [111] axis becomes unstable.
    }
    \label{fig:111magnetic field spectrum}
\end{figure}

\begin{figure}[tbp]
    \centering    \includegraphics[width=\columnwidth]{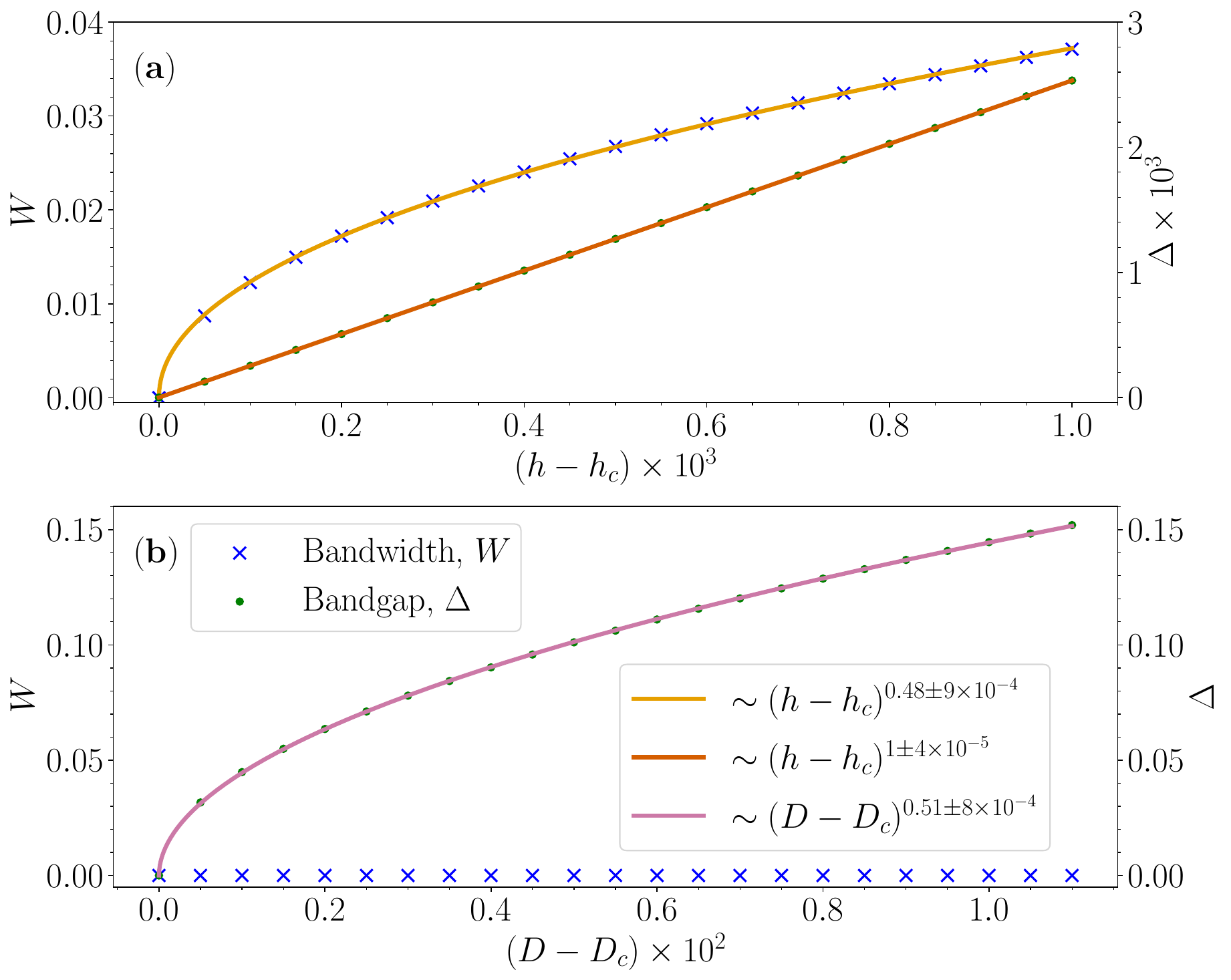}
    \caption{We analyze the scaling of the bandgap and lowest band bandwidth for trivial reference states stabilized by (a) $[111]$ Zeeman field, and 
    (b) $[111]$ easy-plane single-ion anisotropy (SIA). The bandgap scales linearly for the magnetic-field deformation, while 
    it displays an (approximate) square-root behaviour for the SIA. The bandwidth of the lowest band for the reference state stabilized by the
    SIA deformation vanishes, reflecting its flatness in the entire large-$D/J$ phase [see panel (b)], while it exhibits an (approximate) square-root scaling in the field-polarized regime as shown in panel (a).
    }    
    \label{fig:bandgap_pics_both_deformation}
\end{figure}

We first apply this construction to \(H_Q+H_h'\) about the \([111]\)-field-polarized reference state. Diagonalizing \(H^{(2)}\) in momentum space (see Appendix~\ref{app:lfwt} for details), we obtain four linear flavor-wave bands for the two-site unit cell, shown in Fig.~\ref{fig:111magnetic field spectrum}. For sufficiently large \(h/J\), the lowest band is gapped, with gap \(\Delta>0\), and has a unique minimum at the \(\Gamma\) point. Upon decreasing \(h/J\), the gap decreases and eventually closes at
\[
h=h_{\mathrm c}\approx 0.91 J.
\]
At the same time, the bandwidth \(W\) of the lowest band also shrinks and vanishes as \(h\to h_{\mathrm c}\). Extracting both quantities, we find that the numerical data, displayed in Fig.~\ref{fig:bandgap_pics_both_deformation}(a), are well fit by
\[
W\sim (h-h_{\mathrm c})^{1/2},
\qquad
\Delta\sim h-h_{\mathrm c}.
\]
The simultaneous gap closing and band flattening indicate that the high-field polarized state becomes unstable without selecting a unique ordering wavevector. We therefore take this as evidence that, within the mean-field description, no unique ordered state is selected and a large degeneracy of candidate ground states emerges beyond the instability \(h<h_{\mathrm c}\).
Note that the instability is detected at the level of harmonic mean-field fluctuations, and does not, by itself, determine the nature of the state that emerges beyond the instability.

We next perform the same analysis for \(H_Q+H_D'\), now expanding about the large-\(D\) spin-nematic reference state with
\[
\uvec d_i=\frac{1}{\sqrt3}(1,1,1)^\top
\qquad
\text{for all }i.
\]
In this case, the lowest flavor-wave band is perfectly flat, \(W=0\), throughout the regime studied (see also the Supplementary Material~\cite{suppl}). Its gap decreases upon reducing \(D\) and vanishes at
\[
D=D_{\mathrm c}\approx 0.88 J,
\]
with the (approximate) scaling form
\[
\Delta\sim (D-D_{\mathrm c})^{1/2},
\]
as shown in Fig.~\ref{fig:bandgap_pics_both_deformation}(b). Thus, the large-\(D\) nematic product state becomes unstable as the single-ion anisotropy is reduced toward the undeformed quadrupolar model. As in the field-polarized case, the flatness of the soft mode indicates that no unique ordering wavevector is preferred, again pointing to an extensive degeneracy of mean-field ground states for \(D<D_{\mathrm c}\).

This behavior is in clear contrast to more conventional flavor-wave instabilities, such as in the spin-1 bilinear-biquadratic model on the square lattice, where the ferroquadrupolar-to-ferromagnetic transition is signaled by the closing of the flavor-wave gap at a unique ordering wavevector \(\bvec q=(\pi,\pi)^\top\) \cite{pencSpinNematicPhases2011}.

\subsection{Extensive degeneracy of site-factorized mean-field states in the undeformed model} \label{sec:MFproduct}

The LFWT instabilities of the deformed quadrupolar model found in the previous section already point to a large manifold of competing mean-field configurations emerging at, and extending beyond, the corresponding critical deformation strengths.
On their own, however, these instabilities do not establish whether this degeneracy survives all the way to the undeformed limit, nor do they directly reveal the structure of the associated low-energy manifold. 

To address this question, we now analyze the bare quadrupolar Hamiltonian $H_Q$ within a site-factorized variational ansatz.
While this approximation neglects quantum superposition between distinct configurations, it retains the local energetic constraints imposed by the bond-dependent quadrupolar interactions and thus provides a natural framework for identifying the underlying classical manifold selected at the mean-field level. As we show below, this constrained optimization yields an extensive set of degenerate minima, making the origin of the large low-energy manifold explicit.

\subsubsection{Constrained optimization} \label{sec: constr_optm}

To this end, we consider product states
$
\ket{\psi}=\prod_i \ket{d_i}
$
and minimize the variational energy
$
\min_{\ket{\psi}} \braket{\psi|H|\psi}
$.
Because \(\ket{\psi}\) is site-factorized, the expectation value of the quadrupolar Kitaev Hamiltonian factorizes as
\begin{equation}
\label{eq:ProdHquad}
    \langle H_Q \rangle
    =
    \sum_{\langle i,j\rangle,\alpha}
    \langle \tilde Q_i^\alpha\rangle
    \langle \tilde Q_j^\alpha\rangle,
\end{equation}
where, for notational convenience, we introduced $\tilde Q^x \equiv Q^{yz},\tilde Q^y \equiv Q^{zx},\tilde Q^z \equiv Q^{xy}$ with the label \(\alpha=x,y,z\) tied to the corresponding bond type.

Assuming unbroken time-reversal symmetry, the local states \(\ket{d_i}\) may be parametrized by real \(d\)-vectors. Using the identity
$|\langle \vec S\rangle|^2 + |\langle \vec Q\rangle|^2 = 4/3$
(see Appendix~\ref{app:d-vec-formalism}) together with the assumption of vanishing dipolar order, \(\langle \vec S_i\rangle=0\), we obtain the local constraint $|\langle \vec Q_i\rangle|^2 = 4/3$.
We impose this condition with a Lagrange multiplier \(\lambda_i\) on every lattice site,
\begin{equation}
    \braket{H_\lambda}
    =
    -\sum_i \frac{\lambda_i}{2}
    \left(
    \braket{\vec Q_i}^{\,2} - \frac{4}{3}
    \right).
\end{equation}
The variational ground states within the manifold of product states are then obtained from the constrained minimization condition
\begin{equation}
\label{eq:minimize-condition}
    \frac{\partial\bigl(\braket{H_Q}+\braket{H_\lambda}\bigr)}
    {\partial \langle \tilde Q_i^\alpha\rangle}
    = 0.
\end{equation}
This yields the relation
\begin{equation}
\label{eq:Qops_conditions}
    \braket{\tilde Q_{i+\delta_\alpha}^\alpha}
    =
    \lambda_i \braket{\tilde Q_i^\alpha},
\end{equation}
for \(\alpha=x,y,z\), where \(i\) labels a lattice site on either the \(A\) or \(B\) sublattice, and \(\delta_\alpha\) denotes the displacement vector to its nearest neighbor along an \(\alpha\) bond.

For the other two linearly independent quadrupolar operators, which do not appear in \(H_Q\), Eq.~\eqref{eq:minimize-condition} further implies
\begin{equation}
\label{eq:QopsOthertwo}
    \lambda_i \langle Q_i^{x^2-y^2}\rangle
    =
    \lambda_i \langle Q_i^{3z^2-r^2}\rangle
    =
    0.
\end{equation}
Equation~\eqref{eq:Qops_conditions} also imposes a consistency condition on neighboring Lagrange multipliers,
\begin{equation}
\label{eq:lambda_condition}
    \lambda_i \lambda_{i+\delta_\alpha} = 1.
\end{equation}
Using these relations, the extremal energy can be written as
\begin{equation}
\label{eq:E_cl}
    \langle H_Q\rangle
    =
    E_{\mathrm{ext}}
    =
    \sum_{i\in A}
    \lambda_i
    \left(
    \sum_\alpha \langle \tilde Q_i^\alpha\rangle^2
    \right).
\end{equation}
This expression is minimized by choosing \(\lambda_i=-1\) on all sites \(i\in A\), while maximizing \(\sum_\alpha \langle \tilde Q_i^\alpha\rangle^2\). For time-reversal-invariant states with \(\langle \vec S_i\rangle=0\), this maximum is fixed by the local constraint to be \(4/3\).

The resulting variational minima can be constructed explicitly (see Appendix~\ref{sec:AppMeanFieldstates} for details). Each site is restricted to one of four distinguished local states, which we label by the four colors, purple, red, green, and blue:
\begin{align}
\label{eq:4cstates}
    &\ket{\mathrm{p}} = \frac{1}{\sqrt{3}}(1,1,1)^\top,  
    \qquad \ket{\mathrm{b}} = \frac{1}{\sqrt{3}}(1,1,-1)^\top, \nonumber\\
    &\ket{\mathrm{r}} = \frac{1}{\sqrt{3}}(1,-1,1)^\top,  
    \qquad \ket{\mathrm{g}} = \frac{1}{\sqrt{3}}(-1,1,1)^\top .
\end{align}
These four states are normalized but not mutually orthogonal; rather, they form an overcomplete set in the three-dimensional local Hilbert space.
Their quadrupolar expectation values are listed in Tab.~\ref{tab:four_color_Q}.

\begin{table}[ht]
\centering
\begin{tabular}{p{1cm}|p{1cm} p{1cm} p{1cm}}
\hline
state
& $\langle \tilde Q^x\rangle$
& $\langle \tilde Q^y\rangle$
& $\langle \tilde Q^z\rangle$
\\
\hline
$\ket{\mathrm p}$
& $-\frac23$ & $-\frac23$ & $-\frac23$
\\
$\ket{\mathrm b}$
& $-\frac23$ & $-\frac23$ & $+\frac23$
\\
$\ket{\mathrm r}$
& $-\frac23$ & $+\frac23$ & $-\frac23$
\\
$\ket{\mathrm g}$
& $+\frac23$ & $-\frac23$ & $-\frac23$
\\[3pt]
\hline
\end{tabular}
\caption{Expectation values of the off-diagonal quadrupolar operators for the four local color states.}
\label{tab:four_color_Q}
\end{table}
However, neighboring sites cannot be chosen independently. Equations~\eqref{eq:Qops_conditions} and \eqref{eq:lambda_condition} imply the bond constraint
\begin{equation}
\label{eq:minusfourovernine}
    \langle \tilde Q_i^\alpha\rangle
    \langle \tilde Q_{i+\delta_\alpha}^\alpha\rangle
    =
    -\frac{2}{3}\times\frac{2}{3}
    =
    -\frac{4}{9}.
\end{equation}

All product states built from the local states in Eq.~\eqref{eq:4cstates} and satisfying this bond constraint will be referred to as \emph{four-color states}. 
It is useful to depict these states graphically by associating \(\tilde Q^\alpha\) with the \(\alpha\)-type bond emanating from a given site, as illustrated in Fig.~\ref{fig:four color states combined}. In this representation, Eq.~\eqref{eq:minusfourovernine} requires that a ``\(+\)'' and a ``\(-\)'' meet on every bond. The minimization problem is thus reduced to a purely local constraint, and it is immediately clear that the number of admissible mean-field configurations grows extensively with system size. Every such configuration attains the same minimal energy $E_{\min} = -2N/3$,
where $N$ is the number of sites (with $N = 2 N_c$ where $N_c$ is the number of unit cells on the honeycomb lattice). 

A similar four-color structure has previously appeared in the classical Kitaev honeycomb model for dipolar spins \cite{yanClassicalZ22025} and in the Kitaev-Ising spin-orbital model \cite{fontanaSpinOrbital2025}. It is also closely related to the classical dimer-covering manifold underlying the parent construction of the ruby Rydberg spin liquids \cite{verresenUnifyingKitaevMagnets2022}, as well as to the kagome dimer coverings of quantum dimer model \cite{misguich2002quantum}. In the present case, however, the four colors should not be interpreted as distinct microscopic local states, but rather as four distinguished, generally nonorthogonal spin-1 states selected by the constrained mean-field minimization of \(H_Q\). Thus, despite the different local Hilbert spaces, the resulting variational manifold realizes the same four-color configuration space and the same dimer/\(\Ztwo\)-gauge structure.

\subsubsection{Mapping to static \texorpdfstring{$\Ztwo$}{} gauge theory} 

\begin{figure}[tbp] 
     \centering
    \includegraphics[width=\columnwidth]{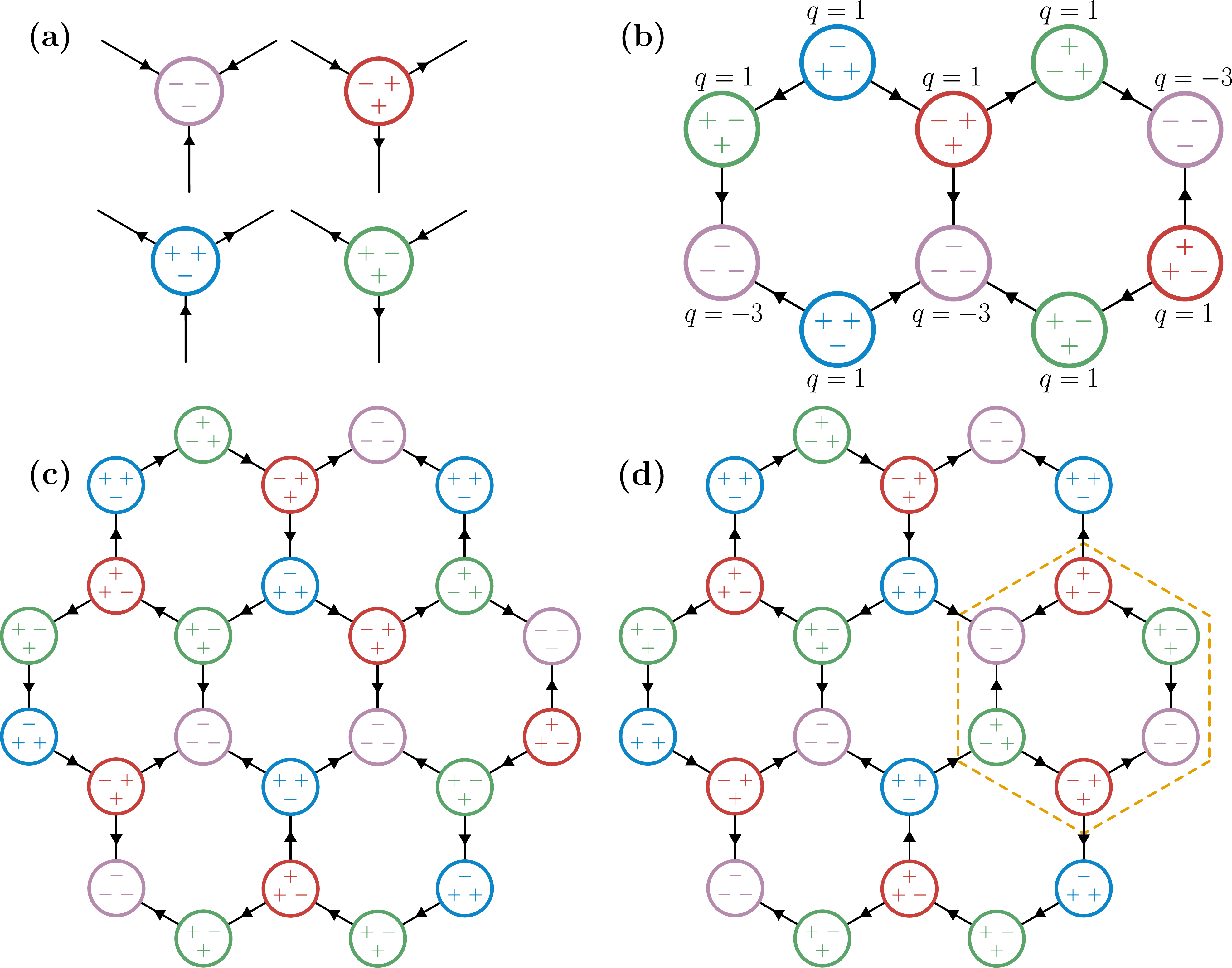}
    \caption{(a) Four color states with associated flux lines to them. (b) The extensive nature of mean field ground states can always be mapped to an emergent gauge theory with charges placed on 
    the lattice and flux lines on the bonds. (c) One representative mean field ground state configuration, (d) a second mean field ground state which differs from (c) at only one hexagon.}
\label{fig:four color states combined}
\end{figure}

The extensive manifold of four-color mean-field minima can be further characterized in terms of a \(\Ztwo\) gauge structure. To this end, on each bond of the honeycomb lattice we draw an arrow pointing from the site carrying a ``\(+\)'' to the one carrying a ``\(-\)'' for the corresponding \(\tilde Q^\alpha\) component, following the arrow representation of kagome dimer coverings \cite{elserKagomeSpin1993}; see also Fig.~\ref{fig:four color states combined}. Each allowed configuration then has a net ``arrow charge'' of \(+1\) or \(-3\) at every vertex. Equivalently, assigning an oriented variable \(\sigma^x_{ij}\in\mathbb Z_2\) to each bond, taken to be positive (negative) when the arrow points from the \(A\) to the \(B\) sublattice (from \(B\) to \(A\)), the mean-field minima obey the local constraint
\[
\prod_{j\in\mathrm{NN}(i)} \sigma^x_{ij}=+1\;(-1)
\qquad\text{for } i\in A\;(B).
\]
This is precisely the odd \(\Ztwo\) Gauss-law constraint on the honeycomb lattice that characterizes kagome dimer coverings \cite{misguich2002quantum,hwangZ2GaugeTheory2015,verresenUnifyingKitaevMagnets2022}. The set of four-color mean-field minima is therefore in one-to-one correspondence with the classical constrained configuration space of kagome dimers. Writing \(\sigma^x_{ij}=\eu^{\iu \pi E_{ij}}\), the constraint may be expressed as
\[
(\nabla\cdot E)_i=q_i \mod 2,
\]
with staggered background charges \(q_i=0\) on the \(A\) sublattice and \(q_i=1\) on the \(B\) sublattice. Within the present mean-field treatment, no dynamics acts within this manifold, which is, therefore, described by a static odd \(\Ztwo\) gauge structure. This identifies the constrained configuration space underlying the isotropic model, but does not by itself establish \(\Ztwo\) topological order in the full quantum problem, which would require quantum dynamics and coherent superpositions within this manifold. Nevertheless, the emergent odd-\(\Ztwo\) gauge structure strongly suggests that quantum fluctuations could naturally lift the extensive classical degeneracy and stabilize a \(\Ztwo\) topologically ordered phase, as obtained in Ref.~\cite{verresenUnifyingKitaevMagnets2022}.

\subsubsection{Topological sectors in the mean-field ground state manifold}

Different topological sectors within the manifold of mean-field ground states can be characterized by loop operators defined as
\begin{equation}
    W_{\mathcal L}
    =
    \prod_{i\in \mathcal L} \eu^{\iu \pi S_i^\alpha},
    \label{eq:W_L}
\end{equation}
where the product runs over sites along a closed loop \(\mathcal L\). Since
\begin{equation}
    [H, W_{\mathcal L}] = 0,
\end{equation}
acting with \(W_{\mathcal L}\) on a given mean-field ground state produces another mean-field ground state by reversing the arrows, namely equivalently exchanging the local \((+,-)\) and \((-,+)\) bond configurations along the closed loop \(\mathcal L\). Since every site on \(\mathcal L\) is incident to two flipped bonds, the local odd-\(\Ztwo\) Gauss-law constraint is preserved. Loop operators therefore generate distinct configurations within the degenerate ground-state manifold and noncontractible loop operators provide a natural way to enumerate the different topological sectors (see also the discussion of related four-color models in Refs.~\cite{pohleSpinNematicsMeet2023,fontanaSpinOrbital2025}).
With periodic boundary conditions, closed loops $\mathcal{L}$ are classified by winding parities $(m,n) \in \mathbb{Z}_2 \times \mathbb{Z}_2$, which specify their winding numbers modulo two.

\subsection{Stability of degenerate mean-field states}

\subsubsection{Field-polarized deformation and single-ion anisotropy} 

The extensive mean-field degeneracy persists up to a finite single-ion anisotropy.
Indeed, evaluating the deformation \(H_D'\) within the four-color manifold, one finds
\begin{equation}
    \braket{H_D'}
    =
    D \sum_i \left(
    \braket{\tilde Q_i^x}
    +
    \braket{\tilde Q_i^y}
    +
    \braket{\tilde Q_i^z}
    + 2
    \right)
    =
    2ND
\end{equation}
Since every four-color state satisfies \(\braket{\tilde Q_i^\alpha}\braket{\tilde Q_j^\alpha}=-4/9\) on each \(\alpha\)-bond, it follows that \(\braket{\tilde Q_i^\alpha}+\braket{\tilde Q_j^\alpha}=0\).
Thus, \(H_D'\) does not lift the degeneracy of the four-color manifold at the mean-field level.

The same reasoning applies to the deformation \(H_h'\) corresponding to a magnetic field along the \([111]\) direction, Eq.~\eqref{eq:deformation h}. Since all four-color mean-field ground states are time-reversal symmetric, one has
$\braket{\vec S_i}=0\, \forall i$
and therefore
\begin{equation}
    \braket{H_h'}=0.
\end{equation}
Thus, the magnetic field does not lift the degeneracy within the four-color manifold at the mean-field level.

\subsubsection{Mean-field configurations for anisotropic couplings \texorpdfstring{$J_z \neq J_x,J_y$}{}}

We now extend the mean-field analysis to anisotropic couplings of the quadrupolar model, as a preparation to the quantum perturbative treatment in Sec.~\ref{sec:anisotropy-limit}. As in the isotropic case, we consider product states
\[
|\psi\rangle=\prod_i |\psi_i\rangle
\]
and minimize the variational energy \(\braket{H_Q}\).

Guided by the isotropic solution, we restrict to an antiferroquadrupolar ansatz on every bond,
\[
\braket{\tilde Q_{i+\delta_\alpha}^{\alpha}}=-\braket{\tilde Q_i^\alpha}.
\]
For positive couplings \(J_\alpha>0\), this is the bondwise energy-minimizing arrangement at the mean-field level. Under this ansatz, the energy reduces to
\begin{equation} \label{eq:h-anisotropic-mean-field}
    \langle H_Q\rangle
    =
    -\sum_{i\in A}
    \left(
    J_z \langle \tilde Q_i^z\rangle^2
    +J_x \langle \tilde Q_i^x\rangle^2
    +J_y \langle \tilde Q_i^y\rangle^2
    \right).
\end{equation}
Writing the real \(d\)-vector as \(\uvec d=(d_x,d_y,d_z)^T\), we use translational invariance to parametrize
\begin{equation} \label{eq:q-uvw}
\langle \tilde Q^x\rangle^2=4vw,\qquad
\langle \tilde Q^y\rangle^2=4uw,\qquad
\langle \tilde Q^z\rangle^2=4uv,
\end{equation}
where $u=d_x^2 \geq 0,v=d_y^2 \geq 0$ and $w=d_z^2 \geq 0$. Minimizing $\langle H_Q\rangle$ then reduces to solving a variational problem subject to the constraint $u+v+w =1$.
 The variational maximum is therefore attained at
\begin{equation} \label{eq:def-qzeig}
    \ket{\Qzeigenone}
    =
    \frac{1}{\sqrt{2}}(1,1,0)^\top,
    \qquad
    \ket{\Qzeigentwo}
    =
    \frac{1}{\sqrt{2}}(1,-1,0)^\top,
\end{equation}
namely the two eigenstates of \(Q^z\), corresponding to directors lying in the \(x\)-\(y\) plane.

With the antiferroquadrupolar bond condition, the mean-field ground states in this parameter regime take the form
\begin{equation}
    |\psi\rangle
    =
    \prod_{\langle ij\rangle_z}
    \ket{\tilde{\sigma}^z_{\langle ij\rangle_z}},
\end{equation}
where each \(z\)-bond is independently in one of the two configurations
\begin{equation}
\ket{\tilde{\sigma}^z_{\langle ij\rangle_z}}
=
\ket{\Qzeigenone_i\,\Qzeigentwo_j},
\qquad
\ket{\Qzeigentwo_i\,\Qzeigenone_j}.
\end{equation}
Thus, for sufficiently anisotropic couplings, the mean-field manifold contains \(2^{N/2}\) degenerate configurations, where \(N\) is the number of honeycomb lattice sites.

\subsubsection{Fluctuations beyond mean-field theory: absence of order-by-disorder}

Given the extensive degeneracy of mean-field ground states, quantum fluctuations beyond mean field are expected to lift this degeneracy and select a quantum ground state. The resulting state could either exhibit symmetry-breaking order, corresponding to an order-by-disorder mechanism, or remain quantum disordered and liquid-like. We can exclude the former possibility, at least at the order considered: real-space perturbation theory \cite{zhitomirskyRealSpacePerturbation2015} shows that the leading fluctuation corrections are identical for all four-color configurations and therefore do not select a symmetry-broken state (see Appendix~\ref{sec:AppRealSpacePert} for details).

At the same time, the four-color states are \emph{overcomplete}: although a local \(S=1\) moment may be assigned one of four colors, only three of these states are linearly independent. This makes a direct effective description of tunneling between different four-color configurations less transparent within the spin-1 Hilbert space, since the corresponding many-body states are nonorthogonal. Thus, while the diagonal fluctuation corrections do not select an ordered state, understanding the remaining off-diagonal dynamics requires a more careful treatment of tunneling within this constrained manifold, which we leave for future work.

\section{Anisotropic limit: quantum-mechanical perturbation theory} \label{sec:anisotropy-limit}

To obtain controlled insight into the quadrupolar Hamiltonian \(H_Q\), we now consider the anisotropic regime \(J_z \gg J_x,J_y\) which was mentioned in Sec.~\ref{sec:solvable_limits}.
Setting first \(J_x=J_y=0\), the unperturbed Hamiltonian is
\[
H^{(0)} = J_z\sum_{\langle ij\rangle_z} Q_i^{xy}Q_j^{xy}.
\]
The operator \(Q^{xy}\) has eigenstates \(\ket{\Qzeigenone}\), \(\ket{\Qzeigentwo}\), and \(\ket{\Qzeigenthree}=(0,0,1)^\top\) with eigenvalues \(-1\), \(1\), and \(0\), respectively. On each \(z\)-bond \(b=\langle ij\rangle_z\), the two states
\[
\ket{\uparrow}_b \equiv \ket{\Qzeigenone_i\,\Qzeigentwo_j},
\qquad
\ket{\downarrow}_b \equiv \ket{\Qzeigentwo_i\,\Qzeigenone_j}
\]
span the degenerate ground-state manifold of \(H^{(0)}\). We therefore identify them as pseudospin states with \(\tilde\sigma_b^z=\pm1\).

\subsection{Degenerate Perturbation Theory}

Turning on \(J_x,J_y \ll J_z\), the perturbation
\begin{equation}
V = J_x \sum_{\langle ij\rangle_x} Q_i^{yz}Q_j^{yz}
   + J_y \sum_{\langle ij\rangle_y} Q_i^{zx}Q_j^{zx}
\end{equation}
lifts this \(2^{N/2}\)-fold degeneracy.
Using degenerate perturbation theory within the pseudospin manifold, one finds that the first three orders contribute only an overall energy shift, while the first nontrivial term appears at fourth order (see App.~\ref{sec:AppAnisoPert}).

At this order, there are three types of virtual processes: those involving only \(x\)-bonds, only \(y\)-bonds, and mixed processes with two \(x\)- and two \(y\)-bond perturbations. All of them act as a pseudospin flip \(\tilde\sigma_b^x\) on a \(z\)-bond, giving
\begin{subequations}\begin{align}
H_{\mathcal O(J_x^4)}^{(4)} &= \frac{7J_x^4}{96J_z^3}\sum_b \tilde\sigma_b^x, \\
H_{\mathcal O(J_y^4)}^{(4)} &= \frac{7J_y^4}{96J_z^3}\sum_b \tilde\sigma_b^x, \\
H_{\mathcal O(J_x^2J_y^2)}^{(4)} &= -\frac{7J_x^2J_y^2}{48J_z^3}\sum_b \tilde\sigma_b^x.
\end{align}\end{subequations}
Combining these contributions yields the effective Hamiltonian
\begin{equation}
H_{\mathrm{eff}}^{(4)}
=
\frac{7}{96J_z^3}(J_x^2-J_y^2)^2
\sum_{b=\langle ij\rangle_z}\tilde\sigma_b^x.
\label{eq:effective_hamiltonian}
\end{equation}

Thus, for \(J_x\neq J_y\), the low-energy theory is simply a transverse field acting on the \(z\)-bond pseudospins, whose ground state is (to leading order) the polarized product state
\begin{equation}
\ket{\psi_0}=\prod_b \ket{+}_b,
\qquad
\ket{+}_b=\frac{1}{\sqrt2}
\bigl(\ket{\Qzeigenone\,\Qzeigentwo}+\ket{\Qzeigentwo\,\Qzeigenone}\bigr).
\end{equation}
The anisotropic regime therefore realizes a trivial phase.
This is analogous to the spin-1 Kitaev model, whose anisotropic limit is also trivial \cite{minakawaQuantumClassicalBehavior2019}, but contrasts with the spin-\(\tfrac12\) Kitaev model, which reduces to the toric code and exhibits \(\mathbb Z_2\) topological order \cite{kitaevAnyonsExactlySolved2006}.

At the symmetric point \(J_x=J_y\), the fourth-order effective Hamiltonian in Eq.~\eqref{eq:effective_hamiltonian} vanishes, so the pseudospin degeneracy is not lifted at this order. We have furthermore checked explicitly that the sixth-order contribution also vanishes. 
Thus, within the present perturbative analysis, no nontrivial splitting arises up to sixth order.

\subsection{Symmetry Constraint at $J_x=J_y$}

In fact, we now argue that for \(J_x=J_y\), single-bond
pseudospin term \(\tilde{\sigma}_b^x\) are forbidden \emph{by symmetry}.
The key discrete symmetry is the $\mathcal{M}_z$ symmetry of a joint spatial reflection about the $z$-bonds, paired with a $\pi/2$ spin rotation about the $z$-axis (see also Sec.~\ref{subsec:symm}). This interchanges $x$ and $y$ bonds and further maps
\begin{equation}
\mathcal{M}_z : Q^{yz} \to - Q^{xz} \quad \text{and} \quad Q^{xz} \to Q^{yz}
\end{equation}
Because this is a symmetry of the microscopic Hamiltonian, the effective low-energy Hamiltonian must be invariant, as well.

As \(\tilde{\sigma}_b^\alpha\) denote the Pauli operator acting on the
two-dimensional low-energy doublet of a \(z\)-bond \(b\), on this doublet, the
projected squared quadrupole operators take the form
\begin{equation}
(Q^{yz})^2 \;\longrightarrow\; \frac12(\mathbf{1}-\tilde{\sigma}_b^x),
\qquad
(Q^{xz})^2 \;\longrightarrow\; \frac12(\mathbf{1}+\tilde{\sigma}_b^x).
\end{equation}
Therefore the $\mathcal{M}_z$ operation maps
\begin{equation}
\tilde{\sigma}_b^x \;\longrightarrow\; -\,\tilde{\sigma}_b^x,
\label{eq:flavorExchangeSigmaX}
\end{equation}
such that a single \(\tilde{\sigma}_b^x\) operator is \emph{odd} under the
\(x\leftrightarrow y\) symmetry.

The low-energy effective Hamiltonian must respect the $\mathcal{M}_z$ symmetry along the $x$-$y$ symmetric line. Since a term proportional to \(\tilde{\sigma}_b^x\) is odd under $\mathcal{M}_z$, it cannot appear with a nonzero coefficient.
This explains the disappearance of the single-bond transverse-field term in the
explicit perturbative expansion.

By contrast, products of two such operators are even:
\begin{equation} \label{eq:two-sigmax-even}
\tilde{\sigma}_{b_1}^x\tilde{\sigma}_{b_2}^x
\;\longrightarrow\;
(-\tilde{\sigma}_{b_1}^x)(-\tilde{\sigma}_{b_2}^x)
=
\tilde{\sigma}_{b_1}^x\tilde{\sigma}_{b_2}^x,
\end{equation}
and thus \(x\leftrightarrow y\) symmetry does not forbid two-bond \(XX\)-type
interactions.

\subsection{Effective Hamiltonian at $J_x=J_y$ up to the 8th order}

\begin{figure}[tbp]
    \centering
    \includegraphics[width=0.7\columnwidth]{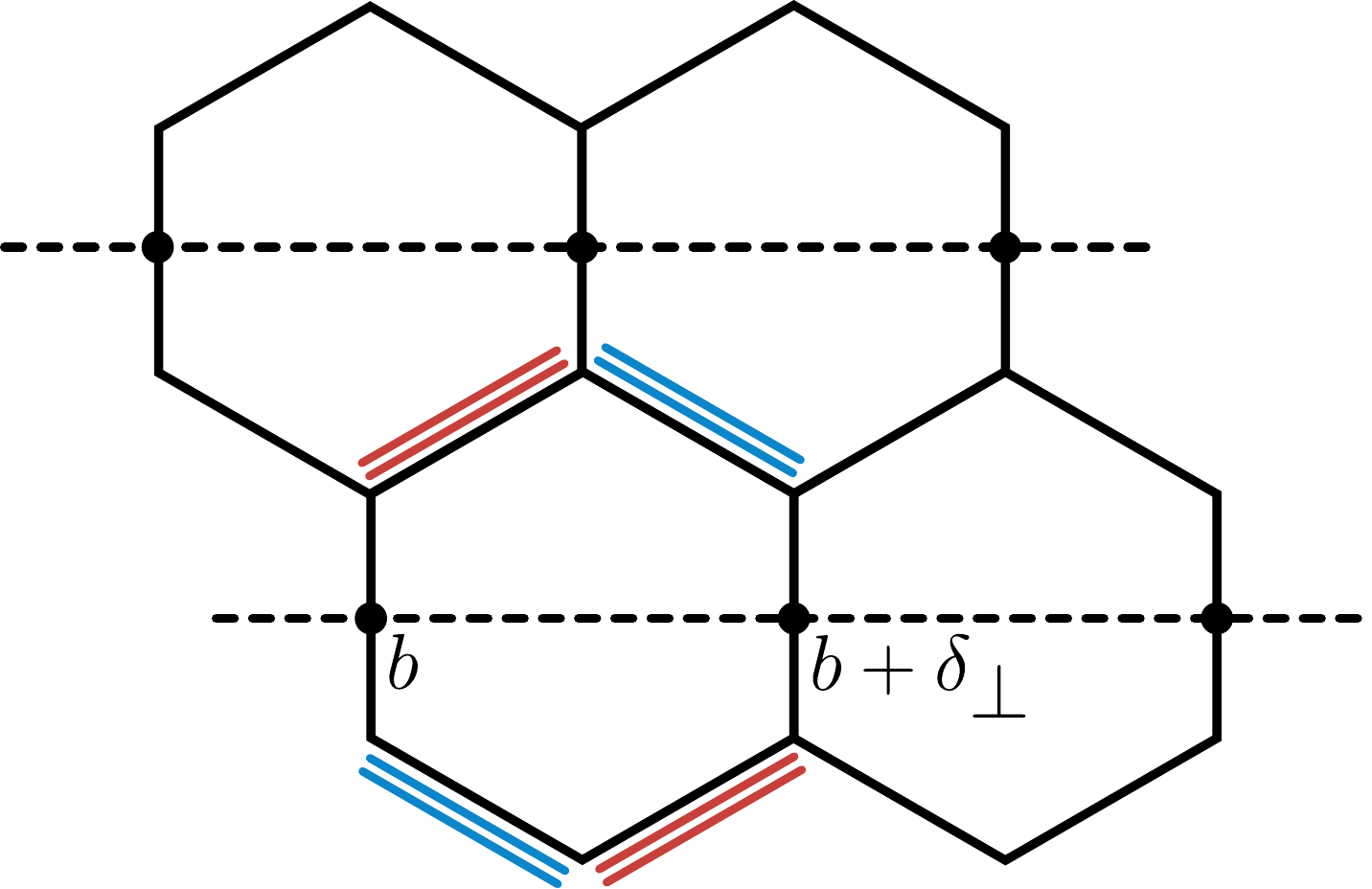}
    \caption{At eighth-order perturbation theory, Ising interactions $\tilde{\sigma}_{b}^x \tilde{\sigma}_{b+\delta_\perp}^x$ emerge between two pseudospins $b,b+\delta_\perp$ associated with two neighboring $z$-bonds (here, red and blue lines indicate the application of the perturbation $ V \sim (J_x/J_z)$ and $J_y/J_z$, respectively). The effective Hamiltonian then realizes Ising chains of pseudospins $K \sum_{b\in \mathrm{chains}} \tilde{\sigma}_b^x\tilde{\sigma}_{b+\delta_\perp}^x$ (as indicated by dashed lines) with $K= \mathcal{O}(J_x^4J_y^4)$, which become coupled only at higher orders in perturbation theory.}
    \label{fig:eight order pt process}
\end{figure}

The combination of the honeycomb lattice geometry and the fact that the perturbation has to act on each bond twice (to yield a non-trivial projection into the ground-state manifold) implies that, for $J_x = J_y$, the symmetry-allowed couplings of the form \eqref{eq:two-sigmax-even} can only emerge at eighth-order perturbation theory (or higher). Indeed, processes as depicted in Fig.~\ref{fig:eight order pt process} yield an effective Ising-type coupling between neighboring $z$-bonds $\tilde{\sigma}_{b}^x \tilde{\sigma}_{b+\delta_\perp}^x$, where $\delta_\perp$ is a lattice vector connecting it to the parallel $z$-bond in the same hexagon.
As a result, at eighth-order perturbation theory in $J_x = J_y \ll J_z$, the effective Hamiltonian decomposes into a sum over decoupled Ising chains for the pseudospin operators associated with $z$-bonds,
\begin{equation}
H_{\mathrm{eff}}^{(8)} = \sum_{\mathrm{chains}} K \sum_{b\in \mathrm{chains}}
\tilde{\sigma}_b^x\tilde{\sigma}_{b+\delta_\perp}^x,
\end{equation}
where each chain has two ground states, given by the two ferromagnetic \(\pm x\)-polarized states, and $K=\mathcal{O}(J_x^4 J_z^4)$.
The microscopic \(\mathcal{M}_z\) symmetry acts in the low-energy \(z\)-bond manifold as \(\tilde{\sigma}_b^x\mapsto -\tilde{\sigma}_b^x\), and is hence realized as a global Ising flip of these pseudospins. The twofold ground-state degeneracy is associated with spontaneously breaking this $\mathbb{Z}_2$ symmetry.

The decoupling of chains implies that $H_{\mathrm{eff}}^{(8)}$ has a ground-state degeneracy of $2^{N_{\mathrm{chains}}}$, which is subextensive (subsystem-type).
We expect that couplings between different chains will be generated at higher orders in perturbation theory and will result in a lifting of this \mbox{(sub-)extensive} degeneracy. However, since no such couplings can emerge at eigth-order perturbation theory, there is a parametrically regime of anisotropies where the model is governed by quasi-1D behaviour. 

\section{Exact parton construction} \label{sec:exact-partons}

Parton constructions are a useful tool for analyzing possible quantum-disordered liquid phases, as they represent the original degrees of freedom in terms of auxiliary variables in which fractionalization becomes manifest.
Such representations generally enlarge the Hilbert space and hence come with an emergent gauge redundancy. While usually approximations are required, in rare cases the associated gauge structure can be formulated exactly in terms of the partons.
Here we construct an representation for the quadrupolar moments in the spin-\(1\) Hilbert space using the parton construction introduced in Ref.~\cite{maZ2HigherSpin2023}, originally formulated for spin-\(S\) Kitaev models, in which the gauge structure can be determined exactly. 
Specifically, we represent a local spin-\(1\) as the triplet sector of two spin-\(\tfrac12\) degrees of freedom, $\vec S=\vec S_1+\vec S_2$ with $\vec S^{\,2}=2$ and then express each spin-1/2 in terms of four Majorana fermions as in the standard Kitaev construction \cite{kitaevAnyonsExactlySolved2006}.

Since two Majorana fermions can be combined into a complex fermion, four Majorana fermions span a four-dimensional Hilbert space. Accordingly, two such sets of four Majorana fermions span a \(16\)-dimensional Hilbert space. To recover the physical three-dimensional local Hilbert space of a spin-\(1\), we impose three constraints.
First, for each \(a=1,2\), we require
\begin{equation}
\label{eq:1st constraint}
\gamma_a^0 \gamma_a^{x} \gamma_a^{y} \gamma_a^{z} = 1,
\end{equation}
which projects each four-Majorana Hilbert space onto a two-dimensional spin-1/2 subspace. We then restrict to the triplet sector of the two spin-1/2 degrees of freedom, corresponding to the condition
\begin{eqnarray}
\label{eq:2nd constraint}
\sum_{\mu=x,y,z} (S^{\mu})^2
&=&
\frac{3}{2}+\frac{1}{2}\sum_{\mu=x,y,z}\sum_{a,b=1}^{2} \gamma_1^0 \gamma_2^0 \gamma_1^{\mu} \gamma_2^\mu 
\nonumber\\
&=&
S(S+1)
=
2.
\end{eqnarray}

These constraints imply a gauge redundancy in the parton representation. To see this, note that the spin operators can be written in bilinear form as
\begin{equation}
S^{\mu}
=
\frac{\mathrm{i}}{2}
\begin{pmatrix}
\gamma_1^0 & \gamma_2^0
\end{pmatrix}
\begin{pmatrix}
\gamma_1^{\mu} \\
\gamma_2^{\mu}
\end{pmatrix},
\end{equation}
which is invariant under
\begin{equation}
(\gamma_1^\mu,\gamma_2^\mu)^\top \to R(\gamma_1^\mu,\gamma_2^\mu)^\top,
\qquad
R\in \mathrm{O}(2).
\end{equation}
Since \(\mathrm{O}(2)=\mathrm{SO}(2)\ltimes \mathbb Z_2\), the gauge structure contains both the continuous \(\mathrm{SO}(2)\) subgroup of proper rotations and a discrete \(\mathbb Z_2\) component associated with improper transformations. Correspondingly, one can define the parton bilinears
\begin{align}
\Gamma^\alpha
=
\frac{\mathrm{i}}{2!}\epsilon_{a_1a_2}\gamma_{a_1}^\alpha \gamma_{a_2}^\alpha
\equiv
\mathrm{i}\gamma_1^\alpha\gamma_2^\alpha,
\qquad
\alpha=0,x,y,z.
\end{align}
These operators are singlets under the \(\mathrm{SO}(2)\) subgroup, but are odd under the \(\mathbb Z_2\) component, namely \(\Gamma^\alpha\to -\Gamma^\alpha\).

Using the Majorana anticommutation relations  $\{\gamma_a^\alpha,\gamma_b^\beta\}=2\delta_{ab}\delta_{\alpha\beta}$,
the quadrupolar operators can then be written as
\begin{equation}
\label{eq:q-parton}
Q^{\alpha\beta}
=
-\mathrm{i}\Gamma^0
\left(
\gamma_1^\alpha\gamma_2^\beta
+
\gamma_1^\beta\gamma_2^\alpha
\right),
\qquad
\alpha,\beta=x,y,z.
\end{equation}
Inserting this into \eqref{eq:Hamil}, we arrive at the parton Hamiltonian
\begin{multline} \label{eq:Hamiltonianparton}
	H = \sum_{(\mu \nu \rho) \in \mathrm{cycl.}(xyz)} \left(-\frac{J_\mu}{4}\right)  \sum_{\langle ij\rangle_\mu} \Gamma_i^0 \Gamma_j^0 \\\times\left(\gamma_{1,i}^\nu\gamma_{2,i}^\rho+\gamma_{1,i}^\rho\gamma_{2,i}^\nu\right) \left(\gamma_{1,j}^\nu\gamma_{2,j}^\rho+\gamma_{1,j}^\rho\gamma_{2,j}^\nu\right),
\end{multline}
where ``$\mathrm{cycl.}(xyz)$'' denotes cylic permutations of bond-type and Majorana flavors.

Following Eq.~\eqref{eq:Hamiltonianparton} and the Majorana anticommutation relations, one finds immediately that \(\Gamma_i^0\) commutes with the Hamiltonian,
\begin{equation}
[H,\Gamma_i^0]=0.
\end{equation}
Hence the \(\Gamma_i^0\) define a set of conserved \(\mathbb Z_2\) variables. In a fixed gauge sector, each \(\Gamma_i^0\) can be replaced by its eigenvalue \(\pm1\), so that the nontrivial dynamics is entirely carried by the remaining Majorana fermions. We therefore omit \(\Gamma_i^0\) in the operator expressions below.

\subsection{Plaquette operators and open string operators in the parton representation}

\begin{figure}[tbp]
\centering
\includegraphics[width=\columnwidth]{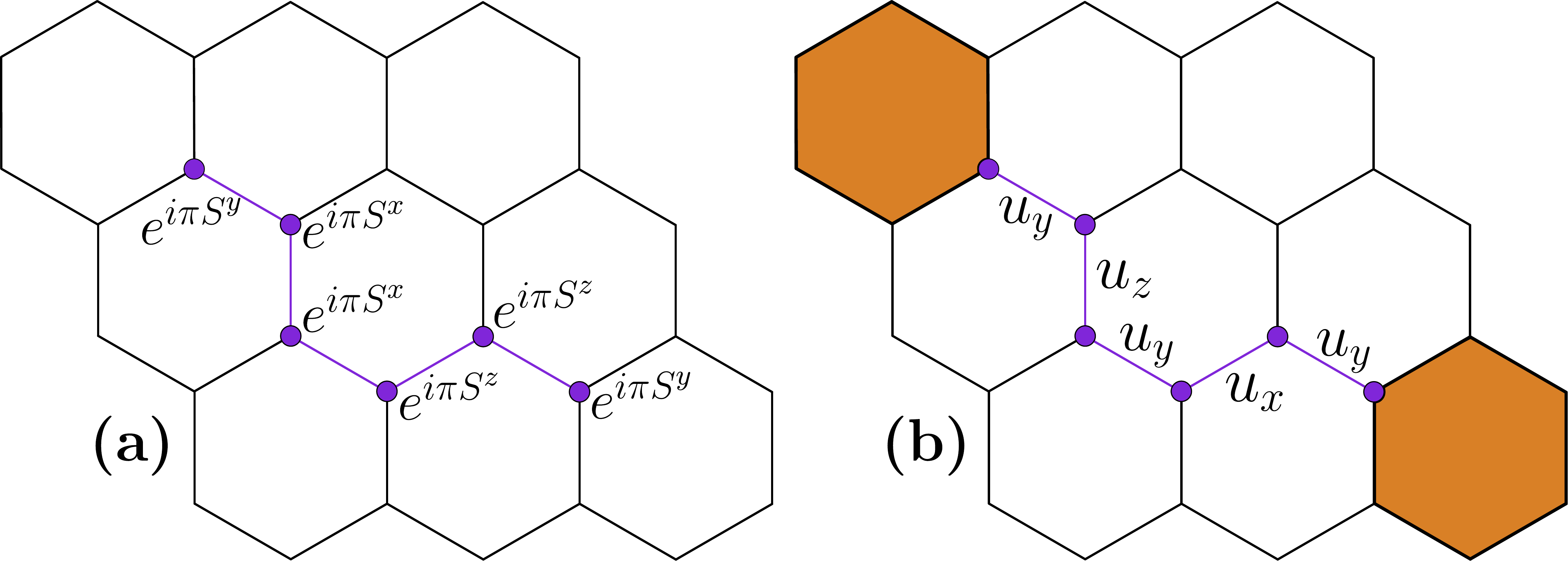}
 \caption{(a) On an open string, we can define a string operator which commutes with the Hamiltonian except at the endpoints, (b) the string operator can then be expressed in terms of gauge fields and 
 endpoint objects. These endpoint excitations are deconfined as they can be moved away from each other without any energy cost. As the plaquette hexagons $W_{\mathcal{L}}$ (indicated in orange) at the two ends of the string operator follow $[W_{\mathcal{L}},u_{ij}^\mu]\neq 0$, flux excitations are also created by the string operator. }
 \label{fig:string_op_decompose}
\end{figure}

As discussed in Sec.~\ref{subsec:symm}, the model has a 1-form symmetry associated with the closed-loop operators \(W_{\mathcal L}\) defined in Eq.~\eqref{eq:W_L}.
The corresponding operator defined on an open string \(\mathcal S\),
\begin{equation}
\mathcal U_{\mathcal S}=\prod_{i\in\mathcal S} \eu^{\iu \pi S_i^{\bar\alpha}},
\end{equation}
creates a pair of excitations at the endpoints. For example,
\begin{equation}
\label{eq:stringop1}
\mathcal U_{\mathcal S}
=
\eu^{\iu\pi S_i^{y}}
\eu^{\iu\pi S_j^{x}}
\eu^{\iu\pi S_k^{x}}
\eu^{\iu\pi S_l^{z}}
\eu^{\iu\pi S_m^{x}}
\eu^{\iu\pi S_n^{x}} 
\end{equation}
as shown in Fig.~\ref{fig:string_op_decompose}(a).

Using \(\eu^{\iu \pi S_j^\alpha}=\Gamma_j^0\Gamma_j^\alpha\), the open-string operator can be rewritten in terms of bond variables and endpoint operators as
\begin{multline}
\label{eq:U_open_gauge}
\mathcal U_{\mathcal S}=
(\mathrm{i}\gamma^y_2\gamma^z_2\gamma^x_1)_i\,
u_{ij}^z u_{jk}^y u_{kl}^x u_{lm}^y u_{mn}^z\,
(\mathrm{i}\gamma^z_1\gamma^x_1\gamma^y_2)_n .
\end{multline}
where we have suppressed the factors of \(\Gamma_j^0\). Here the operators
\begin{equation} \label{eq:exact-u-gauge-field}
u_{ij}^\mu
=
\mathrm{i}
(\gamma_1^\nu\gamma_1^\rho\gamma_2^\mu)_i
(\gamma_2^\nu\gamma_2^\rho\gamma_1^\mu)_j,
\qquad
(\mu\nu\rho)\in \mathrm{cycl.}(xyz),
\end{equation}
are defined on nearest-neighbor bonds \(\langle ij\rangle_\mu\) of type \(\mu\). They mutually commute and also commute with the Hamiltonian,
\begin{equation}
[H,u_{ij}^\mu]=0,
\qquad
[u_{ij}^\mu,u_{kl}^\nu]=0,
\end{equation}
and therefore define a \(\mathbb Z_2\) gauge field. As shown in Figure~\ref{fig:string_op_decompose}, this representation makes it manifest that the string is tensionless, since its interior is built entirely from the bond variables \(u_{ij}^\mu\) and therefore does not cost energy. The only energetic contribution comes from the two endpoint operators. 

For the open string in Eq.~\eqref{eq:U_open_gauge}, the choice of endpoint operator can differ by a local operator. For a given nearest-neighbor bond of type \(\mu\), there are two allowed terminating \(\pi\)-spin rotations, as summarized in Table~\ref{tab:endpoint_operators}. In particular, for a \(z\)-type endpoint at site \(i\), the string may terminate with either \(\eu^{\iu \pi S_i^y}\) or \(\eu^{\iu \pi S_i^x}\), whose parton representatives
\begin{equation}
\mathrm{i}(\gamma_{2}^{y}\gamma_{2}^{z}\gamma_{1}^{x})_i,
\quad \text{and} \quad
\mathrm{i}(\gamma_{2}^{z}\gamma_{2}^{x}\gamma_{1}^{y})_i,
\end{equation}
respectively.
Likewise, at the right endpoint \(n\), one may choose either \(\eu^{\iu \pi S_n^x}\) or \(\eu^{\iu \pi S_n^y}\), with the corresponding parton operators obtained analogously. For an endpoint adjacent to a \(\mu\)-type bond, the two allowed terminating operators have parton representatives whose product, by the local identity constraint, gives the remaining local operator \(\eu^{\iu \pi S^\mu}\).

\begin{table}[t]
\centering
\setlength{\tabcolsep}{3pt}
\renewcommand{\arraystretch}{1.15}
\setlength{\tabcolsep}{3pt}
\begin{tabular}{c c c}
\hline\hline
Nearest Bond  & Physical Endpoint & Parton Representative \\
\hline
\(z\) & \(\eu^{\iu \pi S^x}\) &
\(
\iu\,\gamma_2^z\gamma_2^x\gamma_1^y
=
\iu\,\gamma_1^z\gamma_1^x\gamma_2^y
\)
\\
 & \(\eu^{\iu \pi S^y}\) &
\(
\iu\,\gamma_2^y\gamma_2^z\gamma_1^x
=
\iu\,\gamma_1^y\gamma_1^z\gamma_2^x
\)
\\
\hline
\(x\) & \(\eu^{\iu \pi S^y}\) &
\(
\iu\,\gamma_2^x\gamma_2^y\gamma_1^z
=
\iu\,\gamma_1^x\gamma_1^y\gamma_2^z
\)
\\
 & \(\eu^{\iu \pi S^z}\) &
\(
\iu\,\gamma_2^z\gamma_2^x\gamma_1^y
=
\iu\,\gamma_1^z\gamma_1^x\gamma_2^y
\)
\\
\hline
\(y\) & \(\eu^{\iu \pi S^z}\) &
\(
\iu\,\gamma_2^y\gamma_2^z\gamma_1^x
=
\iu\,\gamma_1^y\gamma_1^z\gamma_2^x
\)
\\
 & \(\eu^{\iu \pi S^x}\) &
\(
\iu\,\gamma_2^x\gamma_2^y\gamma_1^z
=
\iu\,\gamma_1^x\gamma_1^y\gamma_2^z
\)
\\
\hline\hline
\end{tabular}
\caption{Possible endpoint operators for an open string adjacent to a bond of type \(\mu\). The equivalent parton representatives follow from the local identity constraint \(\gamma_1^x\gamma_1^y\gamma_1^z\gamma_2^x\gamma_2^y\gamma_2^z=1\).}
\label{tab:endpoint_operators}
\end{table}

The bond \(u_{ij}^\mu\) and endpoint operators transform nontrivially under the \(\mathrm{O}(2)\) gauge redundancy. Since  \(u_{ij}^\mu\) commutes with the Hamiltonian, it defines static \(\mathbb Z_2\) variables for each energy eigenstate.
This \emph{always} nonzero configuration of \(u_{ij}^\mu\) therefore indicates that the larger \(\mathrm{O}(2)\) gauge structure is Higgsed down to a residual \(\mathbb Z_2\) subgroup.
Correspondingly, the endpoint defects are naturally identified as deconfined excitations carrying \(\mathbb Z_2\) gauge charge.

\subsection{Bosonic endpoint excitation as a fermion-flux bound state}

Besides the endpoint operator itself, the open string also creates a flux excitation. This can be seen from the fact that a piece of the string operator, $\tilde{u}_{ij}^\mu$, fails to commute with the plaquette operator at its endpoints, so that the corresponding plaquette term changes sign and acquires eigenvalue \(-1\).
In other words, an open string creates not only fermionic excitations at endpoints, but also a pair of fluxes at its two ends. The physical endpoint excitation is therefore a bound state of the three-Majorana fermion and a \(\mathbb Z_2\) flux. Since the fermion as \(\mathbb Z_2\) charge has nontrivial mutual braiding with the flux, the resulting bound state is bosonic. 


The physical endpoint excitation is bosonic, so it may in principle condense and thereby drive the system into a trivial phase, as discussed previously in Ref.~\cite{maZ2HigherSpin2023}.
This is consistent with the general result from the 1-form symmetry analysis:
\begin{quote}
\emph{%
For integer spin \(S\in\mathbb Z\), the \(\mathbb Z_2^{[1]}\) 1-form symmetry is anomaly-free, so there is no symmetry obstruction to a symmetric short-range-entangled phase} \cite{liuSymmetriesAnomalies2024}.
\end{quote}
At the same time, however, the endpoint boson may transform nontrivially under microscopic symmetries. In that case, while its condensation removes the topological structure, it can also induce spontaneous symmetry breaking, depending on the symmetry quantum numbers carried by the condensed boson. We therefore next examine the symmetry transformation properties of the bosonic endpoint excitation.

\subsection{Symmetry properties of the bosonic endpoint excitation}

Under the $\mathcal{M}_z$ transformation of joint lattice reflection (about the $z$-bonds) and $\pi/2$-spin rotations around the $z$ axis, the physical endpoint operators are permuted (since $\eu^{\iu \pi S^y} = \eu^{-\iu \pi S^y}$),
\begin{equation}
	\mathcal{M}_z: \eu^{\iu\pi S^x}\leftrightarrow \eu^{\iu\pi S^y}, \qquad \eu^{\iu\pi S^z}\mapsto \eu^{\iu\pi S^z}.
\end{equation}
Similarly, each parton representatives contain one $\gamma^x$ and one $\gamma^y$-Majorana, which upon applying $\mathcal{M}_z$ have to be reordered back to the chosen convention. For example, for a $z$-type endpoint,
\begin{eqnarray}
\mathcal{M}_z: F_1 &=& \iu\,\gamma_2^y\gamma_2^z\gamma_1^x
\;\mapsto\;
\,\iu\,\gamma_2^z\gamma_2^x\gamma_1^y = F_2,
\nonumber\\
\mathcal{M}_z: F_2 &=&\iu\,\gamma_2^z\gamma_2^x\gamma_1^y
\;\mapsto\;
\,\iu\,\gamma_2^y\gamma_2^z\gamma_1^x =  F_1. 
\end{eqnarray}
This symmetry action has direct consequences for the fate of the bosonic endpoint excitations which involve $F_1$ and $F_2$.
Since the boson is coupled to a \(\mathbb Z_2\) gauge field, its condensation can in principle Higgs the remaining \(\mathbb Z_2\) gauge structure that results in a topological trivial phase.
At the same time, because the bosonic endpoint excitation transforms nontrivially under $\mathcal{M}_z$, its condensation can in general be expected to break this symmetry.
The phase in which this bosonic endpoint excitation remains gapped is therefore naturally realizes nontrivial PSG structure \cite{wen02}.
Let \(B_1\) and \(B_2\) denote the two bosonic endpoint operators, obtained as bound states of the corresponding endpoint fermions \(F_1,F_2\) with the endpoint flux. 
Under $\mathcal{M}_z$, the two endpoint species are related by
\begin{equation}
\mathcal{M}_z: B_1 \;\mapsto\; \,B_2,
\qquad
B_2 \;\mapsto\; \,B_1.
\end{equation}
Equivalently, we may define (anti-)symmetric combinations $B_\pm = B_1 \pm B_2$ which transform as
\begin{equation}
\mathcal{M}_z: B_+ \;\mapsto\; \,B_+,
\qquad
B_- \;\mapsto\; \,-B_-.
\end{equation}
Note that the condensation of $B_+$ preserves $\mathcal{M}_z$-symmetry,
$\langle B_+\rangle=\phi \neq 0$, while the condensation of any other superposition (of \(B_+\) and \(B_-\)) breaks $\mathcal{M}_z$. In either case, the condensation of the bosonic operator signals a Higgs transition out of the topologically ordered state into a trivial phase.

\subsection{String diagnostics for boson condensation and trivialization}

For the trivial state obtained by condensing endpoint bosons, the long-distance behavior of the open-string operators is determined by the relative pattern of \(\langle B_1\rangle\) and \(\langle B_2\rangle\).
The open-string operators provide a direct probe of the endpoint-boson sector and hence of possible pathways for obtaining a trivial product state within the parton construction, with or without \(\mathcal{M}_z\) symmetry breaking.
Schematically, an open string on a path \(\mathcal S\) may be written as
\begin{equation}
\mathcal U_{ab}(\mathcal S)= F_a(i)\Big(\prod_{\ell\in\mathcal S}u_\ell\Big)F_b(j),
\qquad a,b=1,2,
\label{eq:open_string_ab}
\end{equation}
where \(F_{1,2}\) denote the two fermionic endpoints and the bulk string is built from the \(\mathbb Z_2\)-valued bond fields \(u_\ell\). Since an open string also creates a \(\mathbb Z_2\) flux \(m\) at each endpoint, the physical localized endpoint excitation is the bosonic bound state
\begin{equation}
B_a \sim F_a \times m.
\end{equation}
Thus Eq.~\eqref{eq:open_string_ab} should be understood as a parton decomposition of a string operator whose physical endpoint defects are bosons.

The string operators diagnose whether the endpoint bosons condense. In a phase without their condensation, the expectation value of a long open string decays to zero. In a gapped phase, this decay is expected to be exponential in the endpoint separation,
\begin{equation}
\langle \mathcal U_{ab}(\mathcal S)\rangle \sim e^{-L/\xi},
\end{equation}
with \(L\) the string length or endpoint separation and \(\xi\) a finite correlation length. By contrast, if the endpoint bosons condense, the long-distance open-string expectation is expected to approach a nonzero constant. More precisely, at asymptotically large separation one expects clustering, so that
\begin{equation}
\langle \mathcal U_{ab}(\mathcal S)\rangle
\sim
\langle B_a\rangle \langle B_b\rangle
\Big\langle \prod_{\ell\in\mathcal S}u_\ell\Big\rangle
\label{eq:string_factorization}
\end{equation}
up to connected corrections that are exponentially suppressed in a gapped phase. In a fixed \(\mathbb Z_2\) flux sector, \(\prod_{\ell\in\mathcal S}u_\ell=\pm1\), so the bulk string contributes only an overall sign to the open-string expectation value.

It is therefore useful to list the expected long-distance signals for the different condensation patterns:

\begin{enumerate}
    \item \textbf{No endpoint-boson condensation:}
    \begin{equation}
    \langle B_a\rangle=0,
    \end{equation}
    Then all long open strings decay to zero,
    \begin{equation}
    \langle \mathcal U_{ab}\rangle
    \longrightarrow 0
    \qquad (L\to\infty).
    \end{equation}
    This is the expected behavior if the endpoint defects remain fully uncondensed, and a deconfined $\mathbb{Z}_2$ gauge structure remains.

    \item \textbf{\(\mathcal{M}_z\)-symmetric condensation:}
    \begin{equation}
    \langle B_+\rangle=\phi\neq 0,\qquad \langle B_-\rangle=0.
    \end{equation}
    Then, $\langle B_1\rangle = \langle B_2 \rangle =\phi/2$
    and hence the long open strings behave as
    \begin{equation}
    \langle \mathcal U_{11}\rangle \sim \frac{\phi^2}{4},\quad
    \langle \mathcal U_{22}\rangle \sim \frac{\phi^2}{4},\quad
    \langle \mathcal U_{12}\rangle \sim \frac{\phi^2}{4}.
    \end{equation}
    Thus all channels of the open string operator are expected to exhibit equal asymptotic positive values, which can be taken as a characteristic string diagnostic of the \(\mathcal{M}_z\)-preserving trivial/Higgs phase.

    \item \textbf{Spontaneous \(\mathcal{M}_z\)-breaking condensation:}
    \begin{equation}
    \langle B_+\rangle=\alpha,\qquad \langle B_-\rangle=\beta,\qquad \alpha\neq 0.
    \end{equation}
    This corresponds to a finite \(\mathcal{M}_z\)-odd component of the condensate.
    Then
    \begin{equation}
    \langle B_1\rangle=\frac{\alpha+\beta}{2},\qquad
    \langle B_2\rangle=\frac{\alpha-\beta}{2},
    \end{equation}
    which yields
    \begin{equation}
    \langle \mathcal U_{11}\rangle \sim \frac{(\alpha+\beta)^2}{2},\quad
    \langle \mathcal U_{22}\rangle \sim \frac{(\alpha-\beta)^2}{2},\quad
    \langle \mathcal U_{12}\rangle \sim \frac{\alpha^2-\beta^2}{2}.
    \end{equation}
    Thus, generically these three string channels are all inequivalent, signaling spontaneous breaking of the \(\mathcal{M}_z\) symmetry in the endpoint sector. In particular, note that there may be a sign difference between the same-endpoint channels and the mixed channel.

    \item \textbf{Pair condensation without single-boson condensation:}
    \begin{equation} 
    \langle B_1\rangle=\langle B_2\rangle=0,
    \qquad
    \langle B_1B_2\rangle\neq 0.
    \end{equation}
    In this case there is no single-endpoint condensate, so the open-string diagnostics
    \begin{equation}
    \langle \mathcal U_{11}\rangle,\;
    \langle \mathcal U_{22}\rangle,\;
    \langle \mathcal U_{12}\rangle
    \longrightarrow 0
    \qquad (L\to\infty)
    \end{equation}
    still vanish at long distance, since they probe single endpoint condensation rather than pair condensation. 
    The appropriate order parameter is instead the local or nonlocal correlator built from the composite operator \(B_1B_2\) (note further that $(B_1)^2, (B_2)^2 \sim \mathds{1}$ are trivial).
    This would indicate condensation of a bound state while the individual endpoint bosons remain uncondensed. 
    Depending on how the composite transforms under the \(\mathcal{M}_z\) symmetry, such a pair condensate may either preserve or break \(\mathcal{M}_z\). In either case, since the elementary endpoint bosons themselves remain uncondensed, the underlying gauge structure can persist and remain deconfined. 
\end{enumerate}

From this viewpoint, the open-string operators furnish a direct diagnostic of the parton mechanism for trivialization. The relative magnitudes of the asymptotic values of \(\mathcal U_{11}\), \(\mathcal U_{22}\), and \(\mathcal U_{12}\) then determine whether the resulting trivial phase preserves or breaks the \(\mathcal{M}_z\) exchange symmetry.

\subsection{String operators in the anisotropic limit \label{sec:string_anisotropic}}

In this subsection we evaluate the expectation value of string operators in the
anisotropic regime $J_z \gg J_x, J_y$ and show how their behavior changes between the generic case
\(J_x\neq J_y\) and the symmetric line \(J_x=J_y\), in line with the preceding analysis.
The basic point is that the string operator acquires a nontrivial projected form within the low-energy \(z\)-bond manifold.

\subsubsection{Low-energy \(z\)-bond manifold.}
For \(J_z\gg J_x,J_y\), the low-energy Hilbert space on each \(z\)-bond
\(b=\langle ij\rangle_z\) is spanned by (recall the definition of the eigenstates of $\tilde{Q}^z \equiv Q^{xy}$ in Eq.~\eqref{eq:def-qzeig})
\begin{equation}
\ket{\uparrow}_b \equiv |\Qzeigenone_i\,\Qzeigentwo_j\rangle,
\qquad
\ket{\downarrow}_b \equiv |\Qzeigentwo_i\,\Qzeigenone_j\rangle.
\end{equation}
We denote by \(P\) the projector onto the tensor product of these local
two-dimensional spaces. 

The single-site \(\pi\)-spin rotations act on \(|\Qzeigenone\rangle\) and
\(|\Qzeigentwo\rangle\) as
\begin{align}
\eu^{\iu \pi S^z}|\Qzeigenone\rangle &= -|\Qzeigenone\rangle,
&
\eu^{\iu \pi S^z}|\Qzeigentwo\rangle &= -|\Qzeigentwo\rangle, \nonumber
\\
\eu^{\iu \pi S^x}|\Qzeigenone\rangle &= -|\Qzeigentwo\rangle,
&
\eu^{\iu \pi S^x}|\Qzeigentwo\rangle &= -|\Qzeigenone\rangle, \nonumber
\\
\eu^{\iu \pi S^y}|\Qzeigenone\rangle &= |\Qzeigentwo\rangle,
&
\eu^{\iu \pi S^y}|\Qzeigentwo\rangle &= |\Qzeigenone\rangle.
\end{align}
It follows that \(\eu^{\iu \pi S^z}\) acts \emph{within} the low-energy subspace, whereas
a single-site \(\eu^{\iu \pi S^x}\) or \(\eu^{\iu \pi S^y}\) takes one out of it:
\begin{equation}
P\,\eu^{\iu \pi S^{x,y}_{i}}\,P = 0
\end{equation}
when \(i\) is one endpoint site of a \(z\)-bond.

\subsubsection{Projection of the plaquette operator.}
As discussed above, the plaquette operator \(W_p\) acts trivially on the top and
bottom \(z\)-bonds and flips the left and right pseudospins. Explicitly,
\begin{equation}
P\,W_p\,P = \tilde{\sigma}_{b_L}^x \tilde{\sigma}_{b_R}^x ,
\label{eq:Wp_projected_here}
\end{equation}
where \(b_L\) and \(b_R\) are the two parallel \(z\)-bonds of the hexagon.
This identity will be the basic input for the string expectation values below.

\subsubsection{String operators in the anisotropic limit.}
In the anisotropic \(z\)-bond regime, the behavior of string operators is controlled by their projection onto the low-energy \(z\)-bond manifold. A generic open string ending with an incomplete \(z\)-bond and carrying an \(x\)- or \(y\)-type endpoint is projected out,
\begin{equation}
P\,\mathcal U_{\mathcal S}\,P=0,
\end{equation}
because the endpoint operator takes the state out of the low-energy subspace. Thus, when \(J_x=J_y\), the nontrivial projected string sector is necessarily even under $\mathcal{M}_z$.

The distinction between \(J_x\neq J_y\) and \(J_x=J_y\) is reflected in the string averages. 
For \(J_x\neq J_y\), the \(\mathcal{M}_z\) symmetry is already explicitly broken at the Hamiltonian level, so that we can no longer classify the condensate as being $\mathcal{M}_z$-even or odd, and a notion of spontaneous symmetry breaking is no longer applicable.
In that case, single-\(x\) strings (composed of $\eu^{\iu \pi S^x}$) already have nonzero expectation values, as expected from the polarized product state stabilized by the effective fourth-order Hamiltonian \eqref{eq:effective_hamiltonian}.

By contrast, for $J_x = J_y$, our leading-order anisotropic effective theory leads to decoupled Ising chains which spontaneously break the \(\mathcal{M}_z\)-symmetry. A single-\(x\)-string operator is odd under this symmetry, so it acquires opposite nonzero expectation values in two symmetry-broken states for each chain, and a double-\(x\)-string is \(\mathcal{M}_z\) even. 
A comparison of the resulting expectation values (for the cases of $J_x\neq J_y$ and $J_x = J_y)$) of representative single- and double-string operators are summarized in Table~\ref{tab:string_expectation_values}.

However, we suggest that the subextensive degeneracy of chain ground states might not necessarily imply that the full system breaks the $\mathcal{M}_z$ symmetry: instead, higher-order contributions to the effective Hamiltonian could establish non-trivial superpositions of the chain ground states, thereby removing the exponentially large degeneracy and possibly leading to a $\mathcal{M}_z$-symmetric ground state.
In this case, the phase for \(J_z \gg J_x=J_y\) may be topologically non-trivial and is more naturally associated with condensation of paired endpoint defects, or equivalently double-endpoint strings, than with a straightforward trivial Higgs phase obtained from single-boson condensation.
A simple Higgs phase would generically proliferate the elementary endpoint boson and wash out the topological nature, leading to a short-range-entangled state.
Indeed, the leading-order extensive degeneracy is not the usual signature of a simple trivial phase, but instead follows from a emergent subsystem \((\mathbb Z_2)^{N_{\mathrm{chains}}}\) symmetry of the decoupled chain Hamiltonian.
While more analysis (including an analysis of higher-order contributions to the effective Hamiltonian, which we leave for further studies) is required, the persistence of the subextensive degeneracy could suggest that the relevant ordering might be better viewed as chain-resolved order, rather than as the result of a uniform boson condensate.

\begin{table}[t]
\centering
\small
\renewcommand{\arraystretch}{1.2}
\setlength{\tabcolsep}{4pt}
\begin{tabular}{p{0.42\linewidth}|p{0.24\linewidth}|p{0.26\linewidth}}
\hline\hline
String operator 
& Quadrupolar, \(J_x\neq J_y\) 
& Quadrupolar, \(J_x=J_y\) \\
\hline
 \(\prod_{i\in\mathcal S} \eu^{\iu \pi S_i^z}\)
with \(N_{\mathcal S}\) sites
&
\( (-1)^{N_{\mathcal S}}\)
&
\((-1)^{N_{\mathcal S}}\)
\\
\hline
\(\prod_{i\in\mathcal S} \eu^{\iu \pi S_i^x}\)
crossing \(n_b\) complete \(z\)-bonds
&
\(1\)
&
\(\langle \tilde\sigma\rangle^{n_b}\)
\\
\hline
\(\prod_{i=1}^{2n_b-1} \eu^{\iu \pi S_i^x} \eu^{\iu \pi S_{2n_b}^y}\) 
&
 \(-1\)
&
\(-\langle \tilde\sigma\rangle^{n_b}\)
\\
\hline
Parallel double \(x\)-string crossing  \(n_b\) complete \(z\)-bonds
&
\(1\)
&
\(1\)
\\
\hline
Plaquette / closed string \(W_p, W_{\mathcal L}\)
&
\(1\)
&
\(1\)
\\
\hline\hline
\end{tabular}
\caption{Expectation values of representative string operators in the anisotropic quadrupolar model. In the \(J_x=J_y\) column, \(\mathcal M_z\)-broken chain-ordered states and its Ising-chain orders are denoted by
\(\langle \tilde\sigma\rangle=\pm1\).  
An \(\mathcal M_z\)-symmetric combination of the two states would average over \(\langle \tilde\sigma\rangle=\pm1\), causing the \(\mathcal M_z\)-odd single-string entries to vanish for odd \(n_b\), while leaving the even-\(n_b\) and double-string entries unchanged.
}
\label{tab:string_expectation_values}
\end{table}

\section{Majorana mean-field theory} \label{sec:mmft}

Despite the extensive number of conserved quantities, the model $H_Q$ in Eq.~\eqref{eq:Hamil} is not integrable (this is in sharp contrast to the S=1/2 Kitaev model, whose Hamiltonian is exceptionally structured so that the parton construction directly leads to the exact solution).
For the model at hand, we therefore turn to a mean-field treatment of the interacting parton Hamiltonian in Eq.~\eqref{eq:Hamiltonianparton}. Such parton mean-field theories often provide a useful qualitative description of quantum spin liquid phases in frustrated magnets and can be a starting point for incorporating quantum fluctuations \cite{savaryQuantumSpinLiquids2016}.

We perform a mean-field decomposition that reduces the interacting parton Hamiltonian in Eq.~\eqref{eq:Hamiltonianparton} to a theory of quadratic Majorana bilinears supplemented by self-consistency conditions \cite{youSpinOrbitMottKHM2012,seifertFractionalizedFermiLiquids2018}.
We further recall that the parton construction faithfully represents the physical Hilbert space only when the local constraints in Eqs.~\eqref{eq:1st constraint} and \eqref{eq:2nd constraint} are satisfied.
Enforcing these local constraints is generally not feasible without substantial numerical effort, so in the present treatment we impose them only on average over the system.

\subsection{Mean-field decoupling}

Note that, at face value, the Hamiltonian in Eq.~\eqref{eq:Hamiltonianparton} consists of 8-fermion interaction terms, resulting in a large number of possible mean-field channels.
Some of these channels can be excluded on symmetry grounds. In particular, we restrict to decouplings which preserve the exact conservation law $\left[H,\Gamma_i^0\right]=0$, where the local operator is $\Gamma_i^0 = \iu \gamma^0_{1,i} \gamma^0_{2,i}$. This immediately forbids bilinears that mix $\gamma^0$-Majoranas with $\gamma^\alpha$-Majoranas (with $\alpha=x,y,z$).
We further assume that all other onsite Majorana bilinears vanish, namely $\langle \iu \gamma_i^{\alpha} \gamma_i^{\beta} \rangle=0$, so that the parton mean-field decomposition of the quadrupole operator in Eq.~\eqref{eq:q-parton} yields $\braket{Q^{\alpha \beta}_i}_\mathrm{MF} = 0$, in contrast to the product-state mean-field approximation used in Sec.~\ref{sec:MFproduct}.
With these assumptions, the interacting parton Hamiltonian in Eq.~\eqref{eq:Hamiltonianparton} is reduced, at the mean-field level, to bilinears of Majorana fermions. Schematically, one has

\begin{multline} \label{eq:int-decoupling}
    \Gamma_i^0 \Gamma_j^0 \gamma_{1,i}^\nu\gamma_{2,i}^\rho \gamma_{1,j}^\lambda\gamma_{2,j}^\delta \to
     \big(  \Gamma_i^0 \braket{\Gamma_j^0} + \braket{ \Gamma_i^0} \Gamma_j^0  \big) \\ \times \big(\braket{\iu \gamma_{1,i}^\nu \gamma_{1,j}^\lambda} \braket{\iu  \gamma_{2,i}^\rho\gamma_{2,j}^\delta}
     - \braket{\iu \gamma_{1,i}^\nu \gamma_{2,i}^\rho} \braket{\iu \gamma_{1,j}^\lambda \gamma_{2,j}^\delta } \big) \\
     + \braket{\Gamma_i^0} \braket{\Gamma_j^0} \big( \iu \gamma_{1,i}^\nu \gamma_{1,j}^\lambda \braket{\iu  \gamma_{2,i}^\rho\gamma_{2,j}^\delta} 
      - \iu \gamma_{1,i}^\nu \gamma_{2,i}^\rho \braket{ \iu \gamma_{1,j}^\lambda \gamma_{2,j}^\delta} +  \dots \big) \\
    + \mathrm{const.},
\end{multline}
where the ``\dots'' in the third line contain all remaining distinct contractions, and ``const.'' denotes contributions to the mean-field ground state energy (see Appendix~\ref{app:partonmft} for details).
We also note that the constraints in Eqs.~\eqref{eq:1st constraint} and \eqref{eq:2nd constraint} are quartic in the Majorana operators.
For the purpose of mean-field theory, however, they can be rewritten in terms of bilinears as
\begin{equation} \label{eq:quadratic-constraint}
    \left(\iu \gamma_a^0 \gamma_a^z + i\gamma_a^x\gamma_a^y\right)=0 \ \text{for} \ a={1,2}, \ \text{and} \ \sum_{\mu\in \{0,x,y,z\}} \Gamma^\mu =0.
\end{equation}
These constraints are readily enforced (on average) for the effective non-interacting fermions.
Accordingly, the fermionic mean-field Hamiltonian can be organized as
\begin{equation} \label{eq:h-mf-pieces}
    H_\mathrm{MF} = H_{0} + H_{xyz}+H_{\lambda} + \mathcal{E},
\end{equation}
where $H_0$ is a bilinear in the $\gamma^0$-Majorana operators and $H_{xyz}$ contains bilinears of $\gamma^\alpha_{a,i}$-operators with \(\alpha=x,y,z\), and $H_\lambda$ contains the Lagrange multipliers enforcing Eq.~\eqref{eq:quadratic-constraint}, and \(\mathcal E\) is the mean-field energy offset.

Finally, we assume that translation symmetries of the honeycomb lattice remain unbroken so that there are onsite mean fields $M_s = \braket{ \Gamma^0_{i,s}}$ for each sublattice $s=A,B$, as well as a set of bond mean fields $w^{\alpha \beta}_{ab}({\delta})$
\begin{equation} \label{eq:def-w}
    w^{\alpha \beta}_{ab}({\delta}) = \langle \iu \gamma^\alpha_{a,(i,A)} \gamma^\beta_{b,(i+\delta,B)} \rangle,
\end{equation}
defined on the three distinct nearest-neighbor bonds $\delta = x,y,z$.
Here \(a,b=1,2\) are \(\mathrm{O}(2)\) indices. 
In addition, translation symmetry leaves \(3\times 2=6\) independent Lagrange multipliers \(\lambda^\alpha_s\), corresponding to the three bilinear constraints \eqref{eq:quadratic-constraint} on each of the two sublattices \(s=A,B\).

\subsection{O(2) gauge redundancy of mean-field parameters} \label{subsec:o2-gauge-mft}

The parton construction introduces an internal \(\mathrm{O}(2)\) gauge redundancy, reflecting the fact that different parton states may correspond to the same physical state. Concretely, the indices \(a,b=1,2\) carried by the bond mean-field parameters
\begin{equation}
    w_{ab}^{\alpha\beta}(\delta)
    =
    \left\langle \iu\,\gamma^\alpha_{a,(i,A)}\gamma^\beta_{b,(i+\delta,B)} \right\rangle
\end{equation}
label the internal \(\mathrm{O}(2)\) gauge doublet rather than physical degrees of freedom. As a result, the matrix structure of \(w_{ab}^{\alpha\beta}(\delta)\) is defined up to \(\mathrm{O}(2)\) gauge transformations acting on these indices.

More explicitly, under an \(\mathrm{O}(2)\) gauge transformation acting on the partons,
\begin{equation}
    \gamma^\mu_{a,(i,s)} \;\to\; \sum_b (R_s)_{ab}\,\gamma^\mu_{b,(i,s)},
    \qquad R_s\in \mathrm{O}(2),
\end{equation}
with \(s=A,B\) denoting the two sublattices, the bond mean fields transform as
\begin{equation}
    w^{\alpha\beta}(\delta)\;\to\; R_A\, w^{\alpha\beta}(\delta)\, R_B^\top,
\end{equation}
where \(w^{\alpha\beta}(\delta)\) is viewed as a \(2\times 2\) matrix in the indices \(a,b\). It is then useful to decompose \(w_{ab}\) into tensor structures according to its transformation under a common \(\mathrm{SO}(2)\subset \mathrm{O}(2)\) rotation acting on both ends of the bond,
\begin{equation}
    w \;\to\; R\,w\,R^\top.
\end{equation}
For a \(2\times 2\) matrix \(w\), the invariant tensors under this conjugation action are \(\mathds{1}\) and \(\mathds{J}\equiv \iu \sigma^y\), while the remaining symmetric traceless part transforms nontrivially. Thus, if one restricts to the \(\mathrm{SO}(2)\)-invariant sector, the most general ansatz takes the form
\begin{equation} \label{eq:w_ab-so2-singlet}
    w_{ab} = \omega_1 \mathds{1}_{ab} + \omega_J \mathds{J}_{ab}.
\end{equation}
This decomposition is useful because it organizes the possible mean-field channels in a symmetry-adapted way.

Since the gauge redundancy is local, the transformations on the two ends of a bond may be chosen independently. In particular, one may adopt different \(\mathrm{O}(2)\) gauge transformations on the \(A\) and \(B\) sublattices. As a result, the self-consistent mean-field equations admit families of apparently distinct solutions, related by relative \(\mathrm{SO}(2)\) rotations or improper \(\mathbb Z_2\), which are nevertheless physically equivalent.

We have explicitly verified that such gauge-equivalent degeneracies occur in the solutions presented below. It is therefore convenient to fix the gauge and work with a particularly simple representative. In the following, we choose an \(\mathrm{O}(2)\)-diagonal ansatz for the bond mean fields,
\begin{equation}
    w_{ab}^{\alpha\beta}(\delta)
    =
    w^{\alpha\beta}(\delta)\,\mathds{1}_{ab}.
\end{equation}
This is merely a gauge choice and does not restrict the physical content of the mean-field ansatz. In particular, it fixes the relative gauge convention between the two sublattices, including the sign choice associated with \(\Gamma_A^0\) and \(\Gamma_B^0\). The actual sign pattern ultimately realized must then be determined self-consistently.

\subsection{Self-consistent mean-field solutions}

The mean-field self-consistency conditions in \eqref{eq:def-w} together with the optimization of the Lagrange multipliers $\lambda_s^\alpha$) can be solved iteratively. In doing so, we exploit the translational symmetry to diagonalize the quadratic fermionic Hamiltonian in momentum space (see Appendix~\ref{app:partonmft}).

The possible mean-field ansätze may be further constrained by requiring the quantum-disordered state to preserve physical symmetries of the model, see also Sec.~\ref{subsec:symm}.
In particular, the $C_6$ symmetry, implemented as a $60^\circ$ lattice rotation together with the cyclic permutation of spin labels $x \rightarrow y \rightarrow z \rightarrow x$, implies that
\begin{align}\label{eq:C3 eq bilinears}
    &w^{yy}(z) = w^{zz}(x) = w^{xx}(y), \quad w^{xx}(z) = w^{yy}(x) = w^{zz}(y), \nonumber \\ 
    &w^{xy}(z) = w^{yz}(x) = w^{zx}(y),
    \quad w^{yx}(z) = w^{zy}(x) = w^{xz}(y).
\end{align}
If, \emph{in addition} to $C_6$, the system preserves a $\mathcal{M}_z$ of a joint lattice reflection (interchanging $x$ and $y$ bonds) and  $\pi/2$-spin rotation, we have further have the constraints
\begin{equation} \label{eq:C2 eq bilinears-equal}
    w^{yy}(z) = w^{xx}(z), \quad
    w^{zz}(x) = w^{yy}(x), \quad
    w^{xx}(y) = w^{zz}(y).
\end{equation}
while all mean-field parameters which hybridize different Majorana flavors must be equal,
\begin{multline} \label{eq:C2 eq bilinears-vanish}
    w\equiv w^{xy}(z) = w^{yz}(x) = w^{zx}(y) \\
     = w^{yx}(z) = w^{zy}(x) = w^{xz}(y)=0.
\end{multline}
following the transformation as
\begin{equation}
    \gamma^\alpha \mapsto s_\alpha\,\gamma^{\sigma(\alpha)},
\end{equation}
and the bond mean fields transform according to
\begin{equation}
    w^{\alpha\beta}(\delta)
    \xrightarrow{\mathcal{M}_z}
    s_\alpha s_\beta\,
    w^{\sigma(\alpha)\sigma(\beta)}\!\left(\sigma(\delta)\right).
\end{equation}
where we denote $\sigma(x)=y,\sigma(y)=x,\sigma(z)=z$
and further $s_x=s_y=+1,s_z=-1.$

\subsubsection{Linearly implemented $\mathcal{M}_z$-symmetric ansatz}

We first solve the mean-field self-consistency equations constrained by \emph{both} $C_6$ and $\mathcal{M}_z$ symmetry.
Up to the gauge redundancy discussed in Sec.~\ref{subsec:o2-gauge-mft}, we find that in the diagonal gauge the independent mean-field parameters reduce to
\begin{equation} \label{eq:c2-constrained-ansatz-values}
	w^{xx}(y) \approx 0.551 \quad \text{and} \quad M_A = M_B = \pm 1,
\end{equation}
with all other mean-field parameters determined by Eqs.~\eqref{eq:C3 eq bilinears}, \eqref{eq:C2 eq bilinears-equal} and Eq.~\eqref{eq:C2 eq bilinears-vanish}.
We further remark that the first two constraints in Eq.~\eqref{eq:quadratic-constraint} are satisfied automatically so that the corresponding Lagrange multipliers vanish, while the third constraint requires a finite Lagrange multiplier $\lambda^3_s \approx \mp 0.138 $ for $s=A,B$  to be satisfied.

\begin{figure}[tbp]
    \centering
    \includegraphics[width=\columnwidth]{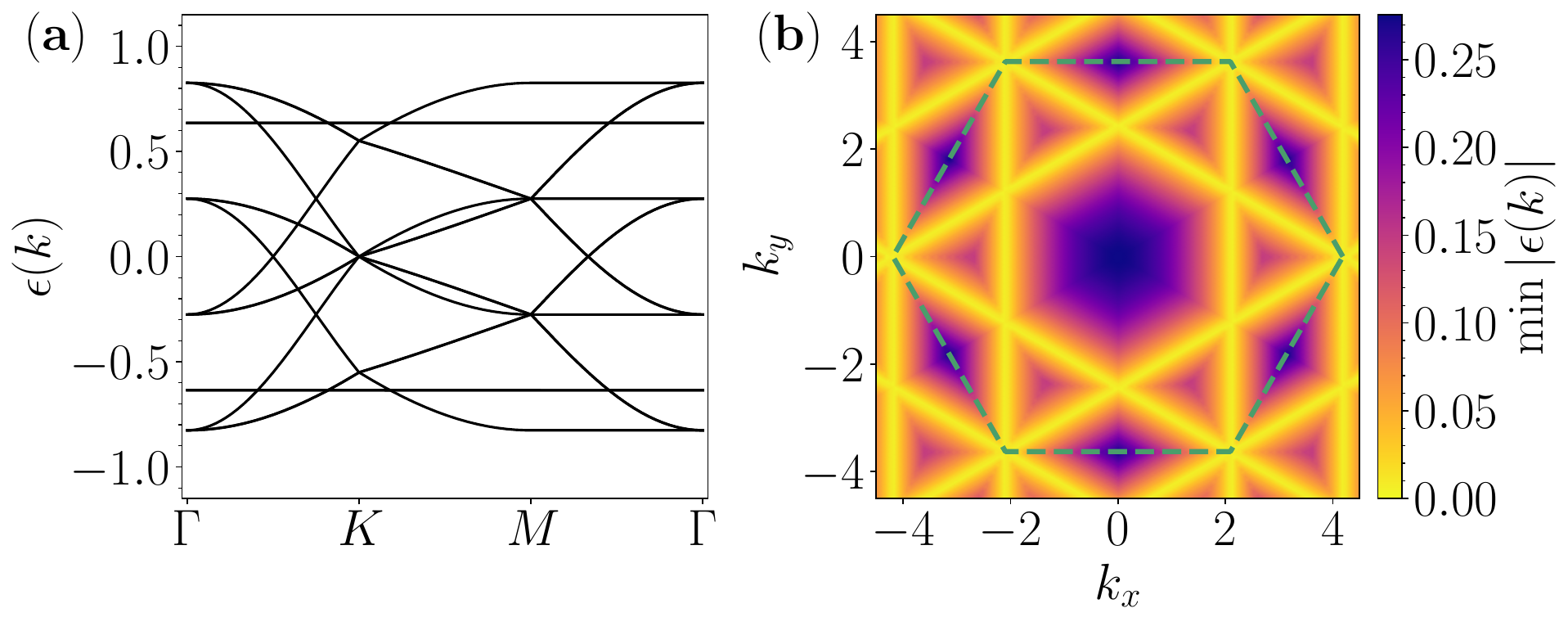}
    \caption{Quasiparticle band structure obtained from Majorana mean-field theory with an ansatz respecting the linearly implemented $\mathcal{M}_z$-symmetry. (a) Cut through band structure along the high-symmetry path $\Gamma$-$K$-$M$-$\Gamma$ (b) Lowest-energy quasiparticle band $\mathrm{min} \, |\epsilon(k)|$ plotted in the first Brillouin zone (indicated in green dashed lines). There are multiple nodal lines in the spectrum, arising from three one-dimensional gapless dispersions of Majorana fermions.}
    \label{fig:min_F_soln_C2enforcedansatz}
\end{figure}

The corresponding Majorana parton band structure is readily obtained from this mean-field ansatz (see Appendix~\ref{app:partonmft}).
We find that it is \emph{gapless} and exhibits a set of nodal lines, as shown in Fig.~\ref{fig:min_F_soln_C2enforcedansatz}.
The origin of this dispersion is transparent from the strongly constrained structure of the ansatz: each \(\gamma^\alpha\)-Majorana disperses only along the two bond types \(\beta\neq \alpha\), so that, for example, \(\gamma^x\) propagates only along the \(y\)- and \(z\)-bonds. The resulting motion therefore takes place along interleaved quasi-one-dimensional chains, and the nodal lines can be understood as arising from one-dimensional Fermi points embedded along three inequivalent directions in momentum space.
Finally, while the finite \(\lambda_s^3\) term hybridizes the two flavors \(\gamma^\alpha_1\) and \(\gamma^\alpha_2\), it does not split their energies, and hence does not alter this nodal-line structure.


\subsubsection{Projectively implemented $\mathcal{M}_z$-symmetric ansatz \label{sec:C2_broken_Topo}}

The linearly $\mathcal{M}_z$-symmetric ansatz discussed above has nodal lines in the parton band structure and may therefore be susceptible to a large number of instabilities when accounting for fluctuations beyond the mean-field approximation.
It also appears to be in tension with the DMRG simulations presented in Ref.~\onlinecite{verresenUnifyingKitaevMagnets2022} which find a \emph{gapped} ground state for the model.
We therefore investigate a less restrictive symmetry implementation at the parton mean-field level.
Specifically, instead of requiring the mean-field parameters to be invariant under the bare action of \(\mathcal M_z\), we allow \(\mathcal M_z\) to be implemented projectively, i.e. up to an internal gauge transformation. In practice, we solve the self-consistency equations imposing the rotational constraints in Eq.~\eqref{eq:C3 eq bilinears}, but not the linear \(\mathcal M_z\) constraints in Eqs.~\eqref{eq:C2 eq bilinears-equal} and \eqref{eq:C2 eq bilinears-vanish}. We then show that the resulting ansatz is nevertheless invariant under \(\mathcal M_z\) once this physical symmetry operation is supplemented by an internal gauge transformation.

We find a self-consistent mean-field solution, up to the gauge redundancies discussed in Sec.~\ref{subsec:o2-gauge-mft}, with values
\begin{multline} \label{eq:c2-broken-ansatz-values}
	w^{xx}(z) \approx \mp 0.614, \quad w^{yy}(z) \approx \pm 0.614 \\
	 \text{and} \quad M_A = -M_B = \mp 1
\end{multline}
while all the flavor-off-diagonal parameters can be taken to vanish (up to numerical accuracy), $w^{xy}(z) \approx 0$.
At first glance, these self-consistent mean-field parameters are not invariant under the bare action of $\mathcal{M}_z$, for example, $\mathcal{M}_z: w^{xx}(z) \to w^{yy}(z)= - w^{xx}(z)$.
However, this extra sign can be compensated by an internal $\SOtwo$ gauge rotation by acting on all Majorana fermions on the $B$ sublattice
\begin{equation}
   \mathcal{G}_B :~ \begin{pmatrix}
        \gamma_{1}^{\mu} \\ \gamma_2^{\mu}
    \end{pmatrix}_B \to  -
    \begin{pmatrix}   
       \gamma_1^{\mu} \\ \gamma_2^{\mu}
    \end{pmatrix}_B.
\end{equation}
This gauge transformation sends every nearest-neighbor \(A\)-\(B\) bond bilinear to its negative,
$w^{\alpha \alpha}(\delta) \to -  w^{\alpha\alpha}(\delta)$ while leaving the onsite $M_A,M_B$ unaffected. 
The combined operation \(\mathcal{G}_B\mathcal M_z\), consisting of the physical \(\mathcal M_z\) transformation followed by this sublattice gauge transformation \(\mathcal{G}_B\), therefore leaves the ansatz invariant.
In this sense, \(\mathcal M_z\) is implemented projectively at the mean-field level.
We leave a full classification of the corresponding projective symmetry group (PSG) to future work \cite{wen02}.

\begin{figure}[tbp]
    \centering
    \includegraphics[width=\columnwidth]{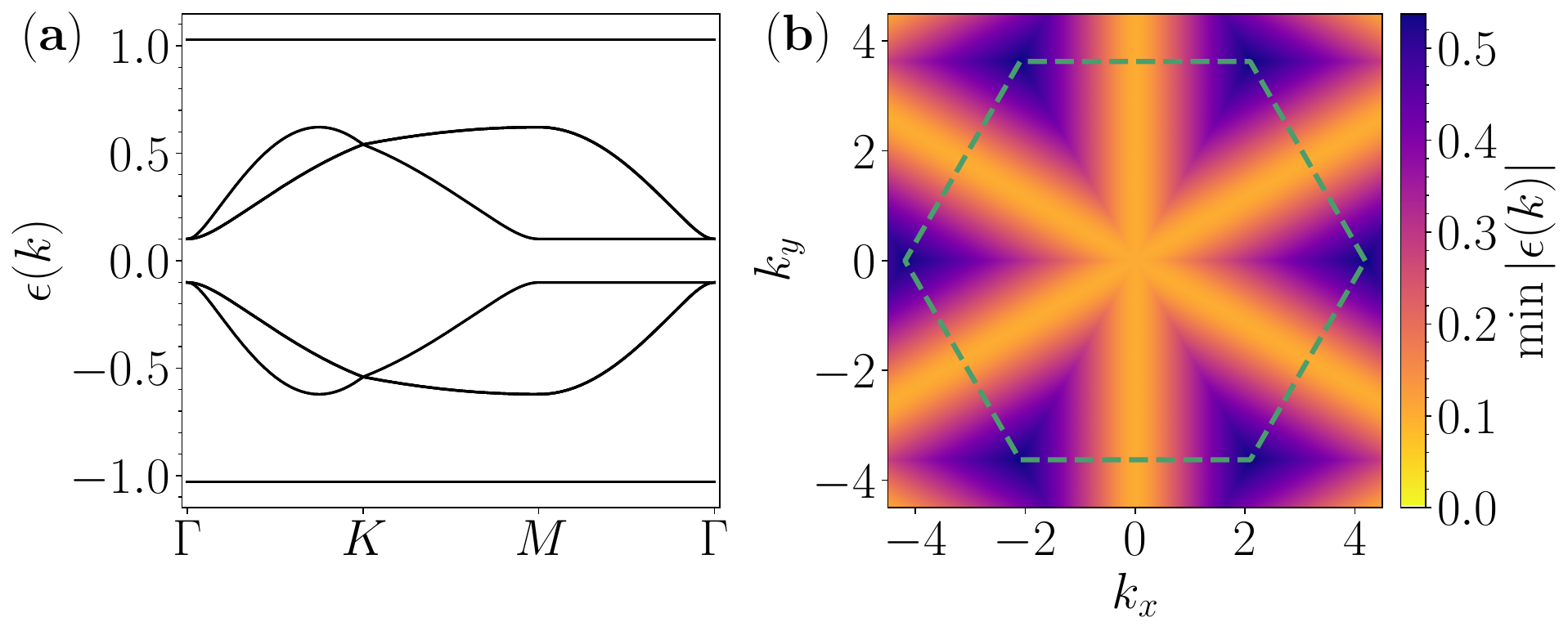}
     \caption{As Fig.~\ref{fig:min_F_soln_C2enforcedansatz}(a) and (b), but now for the self-consistent mean-field saddle point that respects a \emph{projectively implemented} $\mathcal{M}_z$ symmetry.}
    \label{fig:spectrum-c2-broken}
\end{figure}

We find that the parton band structure of this ansatz is fully gapped, as shown in Fig.~\ref{fig:spectrum-c2-broken}.
The origin of the gap can again be understood by viewing the mean-field Hamiltonian as a set of interleaved quasi-one-dimensional Majorana chains.
As in the linearly \(\mathcal M_z\)-symmetric ansatz, each \(\gamma^\alpha\) Majorana disperses only along chains formed by the two bond types \(\beta\neq\alpha\). In the linearly implemented \(\mathcal M_z\) ansatz, these effective chains have a uniform hopping structure and therefore give rise to gapless modes, which appear as nodal lines when embedded in the two-dimensional Brillouin zone.
However, the projective implementation of \(\mathcal M_z\) produces a staggered hopping pattern along these chains.
For example, Eq.~\eqref{eq:c2-broken-ansatz-values}, together with the rotational symmetry constraints, gives
$w^{xx}(y) = -w^{xx}(z)= \pm 0.614$.
Thus, the $x$-Majorana fermions that disperse along chains formed by $y$- and $z$-type bonds experience a \emph{staggered} hopping amplitude.
This unit-cell doubling gaps the one-dimensional modes that would otherwise generate nodal lines in the two-dimensional spectrum.
The same mechanism applies to all three Majorana flavors, yielding a fully gapped parton band structure; an effective one-dimensional tight-binding description is given in the Supplemental Material~\cite{suppl}.
Also note that the spectrum shown in Fig.~\ref{fig:spectrum-c2-broken} contains a gapped flat band which corresponds to the localized $\gamma^0_a$-Majorana fermions, giving rise to the conserved charges $\Gamma^0_{i} = \pm 1$.

We further discuss the plaquette operators $W_{\partial p}$ within the mean-field approximation. The plaquette operator can be written in terms of the 
SO(2)-invariant parton operators as $W_{\partial p} = \prod_{i,\alpha \in \hexagon } \Gamma_{i}^\alpha \Gamma_{i+\alpha}^\alpha$ \cite{maZ2HigherSpin2023}. 
The constraint in \eqref{eq:constraint_bilinear_final_form} together with $M_A = \braket{\Gamma^0_A} = \mp 1 = - M_B = -\braket{\Gamma^0_B}$ 
and $C_6$ symmetry fixes the onsite expectation values
\begin{equation}
	\braket{\Gamma^\alpha_s} = \braket{\iu \gamma^\alpha_{1,(i,s)} \gamma^\alpha_{2,(i,s)}} = \pm (-1)^s \frac{1}{3}
\end{equation}
for $\alpha =x,y,z$.
Here \(s=A,B\) labels the sublattice. Within the present mean-field approximation, we estimate the plaquette operator by factorizing it into onsite \(\Gamma\)-expectation values,
\begin{equation}
    \langle W_{\partial p}\rangle_{\rm MF}
    \simeq
    \prod_{i,\alpha\in \partial p}
    \langle \Gamma_i^\alpha\rangle
    \langle \Gamma_{i+\alpha}^\alpha\rangle .
\end{equation}
This gives
\begin{equation} \label{eq:wp-mft}
    \langle W_{\partial p}\rangle_{\rm MF}
    =
    \left(\frac{1}{3}\right)^6
    \approx 0.0014 .
\end{equation}
We emphasize that within the present decoupling scheme, this estimate uses only the onsite \(\Gamma\)-mean fields. One might try to include additional Wick-type contractions of a factor such as \(\Gamma_i^\alpha\Gamma_{i+\alpha}^\alpha\). However, such terms would require nonzero \(\alpha\)-flavor Majorana bilinears on \(\alpha\)-type bonds. These channels are absent in the mean-field Hamiltonian generated by the present decoupling of Eq.~\eqref{eq:Hamiltonianparton}. Thus Eq.~\eqref{eq:wp-mft} should be understood as the plaquette expectation value within this restricted mean-field scheme.

At the exact level, \(W_{\partial p}\) is an idempotent conserved operator and therefore has eigenvalues \(W_{\partial p}=\pm1\). The non-quantized value in Eq.~\eqref{eq:wp-mft} is therefore not a physical contradiction, but rather a limitation of the mean-field approximation. In particular, the present mean-field treatment is not expected to faithfully resolve nonlocal topological observables such as \(W_{\partial p}\).

\begin{table}[th]
\centering
\begin{tabular}{c|c|c}
\hline
 & Linear \(\mathcal M_z\) & Projective \(\mathcal M_z\) \\
\hline
PSG element
& \(\mathcal M_z\)
& \(G_B\mathcal M_z\) \\
\hline
Gauge transform
& \(G_B=\mathds 1\)
& \(G_B:\gamma^\mu_{a,B}\mapsto-\gamma^\mu_{a,B}\) \\
\hline
\(\mathcal M_z\) action
& \(w^{xx}(z)\mapsto w^{yy}(z)\)
& \(w^{xx}(z)\mapsto -w^{yy}(z)\) \\
\hline
Solution
& \begin{tabular}{@{}c@{}}
\(w^{xx}(y)\approx 0.551\)\\
\(M_A=M_B=\pm1\)
\end{tabular}
& \begin{tabular}{c}
\(w^{xx}(z)\approx\mp0.614\)\\
\(w^{yy}(z)\approx\pm0.614\)\\
\(M_A=-M_B=\mp1\)
\end{tabular} \\
\hline
Effective 1D model
& uniform hopping
& staggered hopping \\
\hline
Spectrum
& gapless; nodal lines
& gapped \\
\hline
\end{tabular}
\caption{
Comparison of the parton mean-field ansatz where \(\mathcal M_z\) is implemented linearly and projectively. In the projective case, \(\mathcal M_z\) is supplemented by a \(B\)-sublattice gauge transformation \(G_B\).
}
\label{tab:linear-projective-psg}
\end{table}

We caution that it is \emph{a priori} unclear to what extent the Majorana mean-field theory can capture aspects of the model's non-trivial gauge structure, in particular in light of the discrepancy between the mean-field parameters $w^{\alpha \beta}(\delta)$ (corresponding to Majorana bilinears) and the $\Ztwo$ gauge field $u_{ij}$ identified in Eq.~\eqref{eq:exact-u-gauge-field}.
However, it is remarkable that the mean-field ansätze lead to a dimensionally reduced (one-dimensional) dispersion of the three Majorana flavors along distinct directions.
We speculate that this might be a precursor to the quasi-1D behaviour realized by the leading-order effective Hamiltonian in the anisotropic limit $J_x = J_y \ll J_z$.
The fact that the projective implementation of the symmetry $\mathcal{M}_z$ (and its $C_6$-conjugate operations) leads to a gapped ground state in the Majorana mean-field theory might be taken as a further indication for the crucial role that this symmetry plays in determining the nature and properties of the system's ground state.

\section{Conclusion and outlook} \label{sec:conclusion}

In this work, we have presented a comprehensive study of the quadrupolar Kitaev model using a range of complementary methods, obtaining controlled insights by perturbing around solvable limits.
Within a semiclassical mean-field approximation, we showed that the model hosts an extensive manifold of ground states characterized by static \(\mathbb Z_2\) variables. Perturbative analyses in the anisotropic limit, combined with insights from a Majorana parton construction, further reveal that the nature of the candidate ground states depends crucially on the mirror symmetry \(\mathcal M\), composed of a real-space lattice mirror operation and a \(\pi/2\) spin rotation. In particular, the relevant gauge charges may carry nontrivial \(\mathcal M\) quantum numbers.
We have detailed different scenarios for the condensation of these gauge charges and concomitant $\mathcal{M}$-symmetry breaking, and the corresponding expected behaviour of open string operators as a diagnostic.
Remarkably, our Majorana mean-field calculations yield Majorana quasiparticle dispersions exhibiting a reduced dimensionality, and we identify a \emph{gapped} state in which the $\mathcal{M}_z$ mirror symmetry is implemented projectively.
We suggest that this state may share features with the observation of a gapped ground state in a previous DMRG study of the model \cite{verresenUnifyingKitaevMagnets2022}.

Several directions remain open for future work.
First, large-scale numerical simulations would be valuable for clarifying the behavior of open-string operators away from the controlled anisotropic limits. In particular, they could determine whether the ground state preserves or breaks the mirror symmetry \(\mathcal M\), and whether the string correlations are consistent with proposed gauge structure.

Second, since the quadrupolar Kitaev Hamiltonian was originally derived as an effective model for the strong-fluctuation regime of ruby-lattice Rydberg quantum arrays, it would be interesting to clarify whether the qualitatively distinct behaviors of open strings with different endpoint operators can be accessed experimentally in these platforms. Rydberg quantum simulators can, in principle, probe string correlators directly~\cite{semeghiniProbingTopologicalSpin2021}, making them a promising setting in which to test the proposed string diagnostics.

Finally, from a materials realization point view, we note that quadrupolar Kitaev interactions could be symmetry-allowed in $S=1$-Kitaev materials. In such systems, they are expected to coexist and compete with more conventional dipolar interactions, including bond-dependent Kitaev exchange and Heisenberg interactions~\cite{stavropoulosMicroscopic2019,mashikoQuantumPhaseTransition2024}. Determining the relative strength of these interactions in candidate materials, and studying their combined phase diagrams quantitatively, will be important for identifying regimes in which a \emph{multipolar liquid} state may be realized.

\begin{acknowledgments}
We gratefully acknowledge discussions with Ruizhi Liu, Achim Rosch and Ruben Verresen. This work is funded by the Deutsche Forschungsgemeinschaft (DFG, German Research Foundation) through SFB 1238, project id 277146847 (UFPS), and the Emmy Noether Program, project id 544397233, SE 3196/2-1 (PS and UFPS). H.M. acknowledges support from the ANR contract associated with the CNRS Chaire Professeur Junior project ``QOQTIB''. H.M. also thanks École Polytechnique for support. Research at the Perimeter Institute is supported in part by the
Government of Canada through Industry Canada, and by the Province of
Ontario through the Ministry of Research and Information.
\end{acknowledgments}


\bibliography{quad_kitaev_vF.bib}


\appendix

\section{$d$-vector formalism for mean-field ground states} \label{app:d-vec-formalism}

To characterize a local spin-1 state, it is convenient to work in the time-reversal-invariant basis
\begin{equation}
    \ket{x}= \iu\frac{\ket{+1}-\ket{-1}}{\sqrt{2}},
    \quad
    \ket{y}= \frac{\ket{+1}+\ket{-1}}{\sqrt{2}},
    \quad
    \ket{z}= -\iu\ket{0},
\end{equation}
where \(\{\ket{+1},\ket{0},\ket{-1}\}\) is the usual \(S^z\)-eigenbasis of the spin-1 Hilbert space. In this basis, $
    \braket{\beta|S^\alpha|\gamma}
    =-\iu\,\epsilon^{\alpha\beta\gamma}$
with $\alpha,\beta,\gamma = x,y,z$,
so that the spin operators can be written as
\begin{equation}
    S^\alpha = -\iu\,\epsilon^{\alpha\beta\gamma}\ket{\beta}\bra{\gamma},
\end{equation}
with summation over repeated indices understood unless stated otherwise. In particular, each basis vector is annihilated by the corresponding spin component, $S^\alpha\ket{\alpha}=0 $ (no summation over \(\alpha\)).
In the same basis, the five quadrupolar operators also take a simple form. For \(\alpha\neq\beta\), the off-diagonal components are
\begin{equation}
    Q^{\alpha\beta}
    =
    -\ket{\alpha}\bra{\beta}
    -
    \ket{\beta}\bra{\alpha},
\end{equation}
while the diagonal traceless components are
\begin{eqnarray}
    Q^{x^2-y^2}
    &=&
    \ket{y}\bra{y}-\ket{x}\bra{x},
    \nonumber\\
    Q^{3z^2-r^2}
    &=&
    \frac{1}{\sqrt{3}}
    \bigl(
        \ket{x}\bra{x}
        +
        \ket{y}\bra{y}
        -
        2\ket{z}\bra{z}
    \bigr).
\end{eqnarray}
In this basis, the quadrupolar operators can be represented as matrices
\begin{equation}
\underline{\underline{Q}}^{xy}
=
-\begin{pmatrix}
0 & 1 & 0\\
1 & 0 & 0\\
0 & 0 & 0
\end{pmatrix},
\,
\underline{\underline{Q}}^{yz}
=
-\begin{pmatrix}
0 & 0 & 0\\
0 & 0 & 1\\
0 & 1 & 0
\end{pmatrix},
\,
\underline{\underline{Q}}^{zx}
=
-\begin{pmatrix}
0 & 0 & 1\\
0 & 0 & 0\\
1 & 0 & 0
\end{pmatrix}\nonumber,
\end{equation}
and
\begin{equation}
\underline{\underline{Q}}^{x^2-y^2}
=
\begin{pmatrix}
-1 & 0 & 0\\
0 & 1 & 0\\
0 & 0 & 0
\end{pmatrix},
\quad
\underline{\underline{Q}}^{3z^2-r^2}
=
\frac{1}{\sqrt{3}}
\begin{pmatrix}
1 & 0 & 0\\
0 & 1 & 0\\
0 & 0 & -2
\end{pmatrix}\nonumber.
\end{equation}
Any local spin-1 state may then be expanded as
\begin{equation}
    \ket{\psi}=\sum_{\alpha=x,y,z} d_\alpha \ket{\alpha},
\end{equation}
and is therefore characterized by a complex three-component vector
\(\uvec d=(d_x,d_y,d_z)^\top\in\mathbb C^3\), defined up to an overall phase. Below, for any three-component vector \(\uvec q\), we use the shorthand \(\uvec q=(q_x,q_y,q_z)^\top\).
Normalization of \(\ket{\psi}\) implies
\begin{equation}
    \uvec d^{\,\dagger}\uvec d = 1.
\end{equation}
Writing $\uvec d=\uvec u+i\uvec v$ with $\uvec u,\uvec v\in\mathbb R^3$,
the spin expectation value takes the form
\begin{equation}
    \langle \vec S \rangle
    \equiv
    \bigl(\langle S^x\rangle,\langle S^y\rangle,\langle S^z\rangle\bigr)^\top
    =
    \bra d \vec S \ket d
    =
    2\,\uvec u\times \uvec v .
\end{equation}
Thus, a nonzero dipolar moment originates from the \emph{relative} phase structure of components of the \(d\)-vector.

The expectation values of the quadrupolar operators defined in Eq.~\eqref{eq:Qopdefinition} are
\begin{equation}
    \bra d Q^{\alpha\beta}\ket d
    =
    \frac{2}{3}\delta_{\alpha\beta}
    -u_\alpha u_\beta
    -v_\alpha v_\beta .
\end{equation}
Equivalently, the dipolar and quadrupolar moments are not independent, but satisfy
\begin{equation}
    \bigl|\bra d \vec S \ket d\bigr|^2
    +
    \bigl|\bra d \mathbf Q \ket d\bigr|^2
    =
    \frac{4}{3},
\end{equation}
where \(\mathbf Q\) denotes the five-component quadrupolar vector introduced in Eq.~\eqref{eq:Qopdefinition}.

This representation is particularly useful for characterizing spin-nematic states. If the dipolar moment vanishes, meaning $\langle S^\alpha\rangle = \bra d S^\alpha \ket d = 0$ for all \(\alpha\),
then $\uvec u\times \uvec v = 0$,
so that \(\uvec u\) and \(\uvec v\) are parallel. It follows that \(\uvec d\) can be chosen real up to an overall \(\mathrm{U}(1)\) phase. In this case, \(\uvec d\) defines a nematic axis rather than an oriented spin vector. Indeed, for real \(\uvec d\) one finds
\begin{equation}
    (\uvec d\cdot \vec S)\ket d = 0,
\end{equation}
which shows that spin fluctuations vanish along the axis specified by \(\uvec d\). A real \(d\)-vector therefore describes a spin-nematic, or quadrupolar, state: it has vanishing spin polarization,
but retains anisotropic spin fluctuations transverse to a distinguished axis, thereby breaking spin-rotation symmetry without developing a dipolar moment.

\section{Linear flavor wave theory} \label{app:lfwt}

In order to study the LFWT spectrum, we choose a local reference basis in which the $\uvec{d}$-vector has coordinates $\uvec{d}' = (0,0,1)^\top$. This is achieved by a local unitary transformation $\uvec{\uvec{U}}=(\uvec{u},\uvec{v},\uvec{d})$, whose columns form an orthonormal basis of $\mathbb{C}^3$, so that $U^\dag U=I$. $\uvec{d}$ is a unit vector in the original basis.
We then choose a unit vector $\uvec{u}$ orthogonal to $\uvec{d}$, and define the third basis vector $\uvec{v} $ orthogonal to both $\uvec{u}$ and $\uvec{d}$. For real $\uvec d$, one convenient choice is $\uvec v =(\uvec u \times \uvec d)^\ast$ \cite{seifertPhaseDiagramsExcitations2022}.

The excitation spectra are obtained from the quadratic-boson Hamiltonian $H^{(2)}$ in Eq.~\eqref{eq:lfwt-expansion}.
Exploiting translational invariance we implement Fourier decompositions for bosonic operators on the $A$ and $B$ sublattices,
$b_{i,\mu}
=
\frac{1}{\sqrt N}
\sum_{\bvec k \in \mathrm{1.\,BZ}}
e^{i\bvec k\cdot \bvec r_i}\,
b_{\bvec k,A}^{\mu}, \quad \quad b_{j,\mu} = \frac{1}{\sqrt{N}} \sum_{\bvec{k} \in 1.\,\mathrm{BZ}} \eu^{\iu \bvec{k} \cdot (\bvec{r}_i +\bvec{\delta})} b_{\bvec{k},B}^{\mu}$.
where \(\bvec\delta\) connects the two sublattices within the unit cell.
In momentum space, the quadratic Hamiltonian takes the Bogoliubov form
\begin{equation}
	H^{(2)} = \frac{1}{2} \sum_{\bvec{k} \in 1.BZ}\psi_{\bvec{k}}^{\dagger} H_{\bvec{k}} \psi_{\bvec{k}},
\end{equation} with Nambu spinor
\begin{equation}
    \psi_{\mathbf{k}} = \left(
    \psi_{\bvec k , A},
    \psi_{\bvec k , B},
    \psi_{-\bvec k , A}^\dag,
    \psi_{-\bvec k , B}^\dag\right)^\top.
\end{equation}
where $\psi_{\bvec k, s}= (
    b_{\bvec k,s}^{x},
    b_{\bvec k,s}^{y})^\top$ for $s=A,B$ sublattices.
$H_{\bvec k}$ is a $8 \times 8$ bosonic BdG Hamiltonian with block form
\begin{eqnarray}
H_{\bvec k}= \left(\begin{array}{cc}
    A_{\bvec k} &  B_{\bvec k} \\
     B_{-\bvec k}^\ast &  A_{-\bvec k}^T
\end{array}\right).
\end{eqnarray}
The eigenmodes are obtained by carrying out a bosonic Bogoliubov transformation \cite{colpaDiagonalizationQuadratic1978}. Concretely, one diagonalizes the matrix \(\Sigma H_{\bvec k}\), where 
\begin{equation}
    \Sigma = \begin{pmatrix}
    \bvec{1}_{4} &  0 \\ 
    0 & -\bvec{1}_{4}
    \end{pmatrix},
\end{equation}
yielding the spectrum $\{\varepsilon_1(\bvec{k}),\dots,\varepsilon_4(\bvec{k}), - \varepsilon_1(\bvec{k}), \dots, -\varepsilon_4(\bvec{k})\}$ with $\varepsilon_i(\bvec{k})  \geq 0$ for $i=1,\dots,4$. These positive eigenvalues give the dispersions of the four bosonic flavor-wave eigenmodes.

\section{Derivation of four-color states}
\label{sec:AppMeanFieldstates}

\subsection{Constrained minimization}

The energy in Eq.~\eqref{eq:E_cl} is straightforwardly minimized when $|\langle \vec Q_i\rangle|^2=\frac{4}{3}$ and $\lambda=-1$. At the same time, the two quadrupolar components not appearing in \(H_Q\) vanish,
$
\langle Q^{x^2-y^2}\rangle=\langle Q^{3z^2-r^2}\rangle=0
$.
For a real \(d\)-vector \(\ket d=(d_x,d_y,d_z)^T\), these conditions read
\[
d_x^2-d_y^2=0,
\qquad
2d_z^2-d_x^2-d_y^2=0,
\qquad
d_x^2+d_y^2+d_z^2=1.
\]
They imply
\[
d_x^2=d_y^2=d_z^2=\frac13,
\]
so the local minimizing states are
\[
\ket d=\frac1{\sqrt3}(\Theta_x 1,\Theta_y 1,\Theta_z 1)^T,
\]
where $\Theta_x, \Theta_y, \Theta_z \in \{+1,-1\}$, leaving \emph{a priori} $2^3 = 8$ possible states. However, note that an overall phase is physically irrelevant, so \(\ket d\) and \(e^{i\phi}\ket d\) represent the same state. 
Identifying \(\ket d\sim -\ket d\), which represent the same spin-nematic state, leaves four distinct local states, namely the four color states in Eq.~\eqref{eq:4cstates}.

%
%

\subsection{Variational problem for anisotropic couplings}

With Eq.~\eqref{eq:q-uvw}, the minimization of the Hamiltonian  \eqref{eq:h-anisotropic-mean-field} maps onto the constrained maxmization problem
\begin{equation}
    f(u,v,w)=4\bigl(J_zuv+J_xvw+J_ywu\bigr) - \frac{\mu}{4}(u+v+w-1), 
\end{equation}
with $\mu$ denoting a Lagrange multiplier to enforce $u+v+w=1$.
An interior stationary point satisfies
\begin{equation}
    J_zv+J_yw=\mu,\quad
    J_zu+J_xw=\mu,\quad
    J_xv+J_yu=\mu,
\end{equation}
which can be solved to yield
\begin{equation}
    u=\frac{J_x+J_z-J_y}{2J_x},\quad
    v=\frac{J_y+J_z-J_x}{2J_y},\quad
    w=\frac{J_x+J_y-J_z}{2J_z}.
\end{equation}
In the regime \(J_x+J_y\le J_z\), one has \(w\le0\), so the maximum must lie on the boundary \(w=0\), i.e. \(d_z=0\). Then \(u+v=1\) and
\begin{equation}
    f=4J_zuv\le J_z,
\end{equation}
with equality at \(u=v=1/2\).

\section{Real-space perturbation theory for mean-field states} \label{sec:AppRealSpacePert}

A real-space perturbation-theory treatment of fluctuations \cite{zhitomirskyRealSpacePerturbation2015} about the four-color mean-field manifold starts from the decomposition
\[
\hat Q_i^a = Q_i^a + \delta \hat Q_i^a,
\qquad
Q_i^a \equiv \langle \hat Q_i^a\rangle ,
\]
where \(Q_i^a\) denotes the expectation value of the quadrupolar operator \(\hat Q_i^a\) in a given four-color state \(\ket{\{Q\}}\). Substituting this into the Hamiltonian gives
\begin{align}
H
&=
\sum_{ij} J_{ij}
\left(
Q_i^a Q_j^a
+
\delta \hat Q_i^a Q_j^a
+
Q_i^a \delta \hat Q_j^a
+
\delta \hat Q_i^a \delta \hat Q_j^a
\right)
\nonumber\\
&=
\underbrace{\sum_{ij} J_{ij} Q_i^a Q_j^a}_{E_{\mathrm{MF}}[\{Q\}]}
+
\underbrace{\sum_{ij} J_{ij}
\Bigl[
Q_i^a(\hat Q_j^a-Q_j^a)
+
(\hat Q_i^a-Q_i^a)Q_j^a
\Bigr]}_{H_0[\{Q\}]}
\nonumber\\
&\hspace{2cm}
+
\underbrace{\sum_{ij} J_{ij}
(\hat Q_i^a-Q_i^a)(\hat Q_j^a-Q_j^a)}_{V[\{Q\}]}\, .
\label{eq:realspaceexpansion}
\end{align}
Here, \(E_{\mathrm{MF}}[\{Q\}]\in\mathbb R\) is the mean-field energy of the configuration \(\{Q\}\). \(H_0[\{Q\}]\) is a noninteracting mean-field Hamiltonian which has the zero-energy state \(\ket{\{Q\}}\). By construction,
\[
E_0[\{Q\}]
=
\bra{\{Q\}} H_0 \ket{\{Q\}}
=
0.
\]
And \(V[\{Q\}]\) contains the residual fluctuation terms.
One may then treat \(V[\{Q\}]\) perturbatively to compute fluctuation-induced corrections to the energy of each four-color configuration. 
To exclude an order-by-disorder mechanism, it is enough to show that the diagonal fluctuation corrections are invariant for all configurations related by loop updates, since such updates generate the four-color manifold within each connected sector. Let \(\ket{\{Q'\}}=W_{\mathcal L}\ket{\{Q\}}\), where \(W_{\mathcal L}\) is a closed-loop operator. Then \(\ket{\{Q'\}}\) is another four-color state with the same mean-field energy,
\[
E_{\mathrm{MF}}[\{Q\}] = E_{\mathrm{MF}}[\{Q'\}] .
\]
Under \(W_{\mathcal L}\), the relevant quadrupolar expectation values change sign along the loop, producing the new configuration \(\{Q'\}\). Accordingly,
\[
H_0[\{Q\}] \to H_0[\{Q'\}],
\qquad
V[\{Q\}] \to V[\{Q'\}].
\]
The \(n\)-th order diagonal correction may be written schematically as
\begin{equation}
E_n[\{Q\}]
=
\bra{\{Q\}}
V
\left(
\frac{1}{E_0-H_0}\,V
\right)^{n-1}
\ket{\{Q\}} .
\end{equation}
It follows that
\[
E_n[\{Q\}] = E_n[\{Q'\}]
\]
for any two loop-related four-color configurations. Hence quantum fluctuations do not energetically distinguish between different four-color states to all orders, and no order-by-disorder selection occurs.

\section{Degenerate perturbation in the anisotropic limit via diagrammic approach} \label{sec:AppAnisoPert}
\begin{figure}[tbp]
    \centering
    \includegraphics[width=\columnwidth]{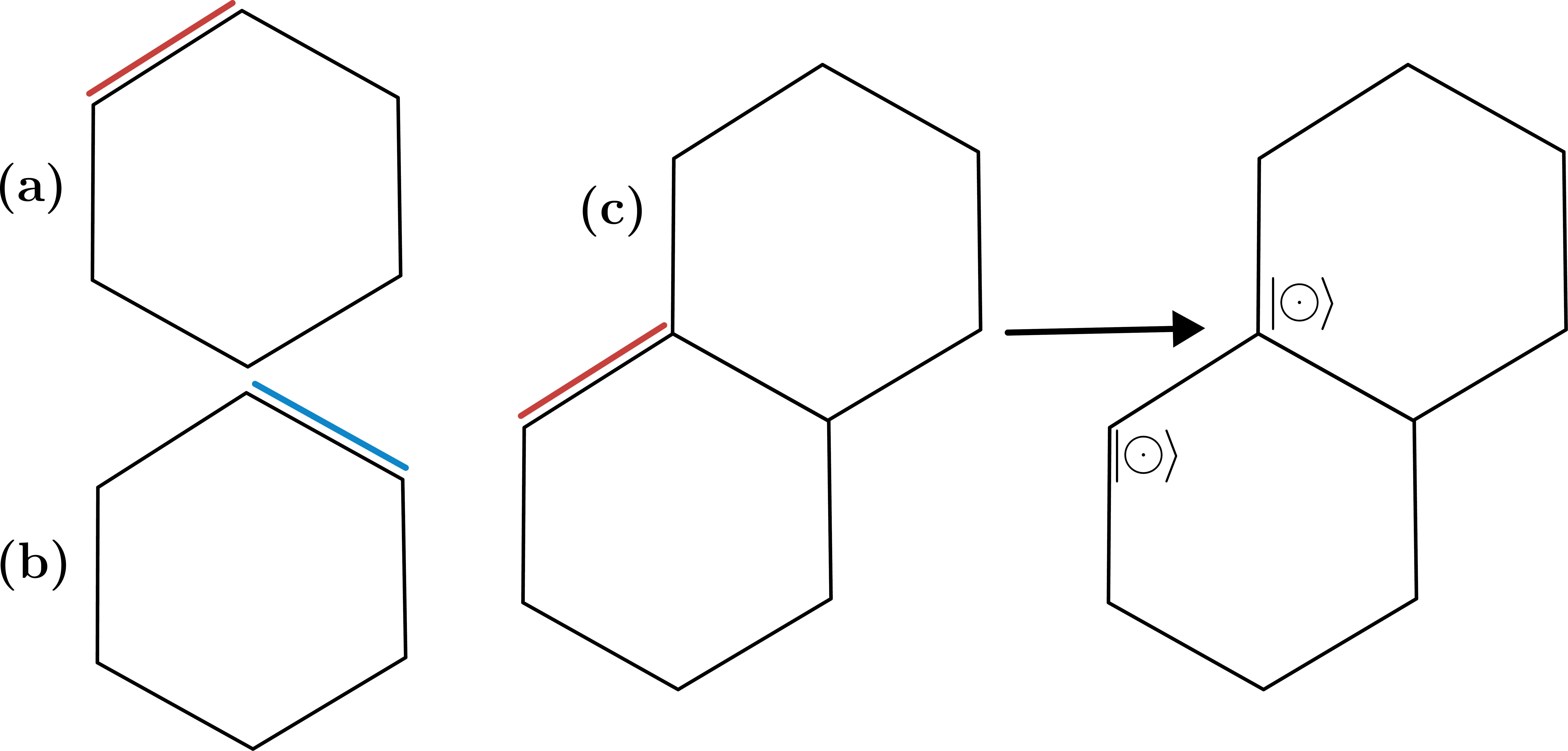}
    \caption{First-order perturbation theory process in graphical notation, denoted by a (a) red line $\left(\mathcal{O}(J_x)\right)$ and (b) blue line $\left(\mathcal{O}(J_y)\right)$, (c) applying a single perturbation on a bond yields states that lie outside the system's low-energy subspace.}
    \label{fig:first_order_pt_combined}
\end{figure}

To derive the effective Hamiltonian within the degenerate \(z\)-bond ground-state manifold, it is convenient to introduce a graphical notation that keeps track of where the perturbation acts: we denote an insertion of \(V_x\) (\(V_y\)) on an \(x\)-bond (\(y\)-bond) by a thick blue (red) line, as illustrated in Fig.~\ref{fig:first_order_pt_combined}. This provides a simple tool for enumerating the virtual processes that contribute in degenerate perturbation theory.

With this notation, one immediately finds that the first-order contribution vanishes,
\[
H^{(1)} = P^\dagger V P = 0.
\]
Explicitly, the perturbing operators \(Q^{zx}\) and \(Q^{yz}\) connect the two low-energy eigenstates of \(Q^{xy}\) to the excited state \(\ket{\Qzeigenthree}\):
\begin{align}
Q^{zx}\ket{\Qzeigenone} &\propto \ket{\Qzeigenthree}, &
Q^{zx}\ket{\Qzeigentwo} &\propto \ket{\Qzeigenthree}, \\
Q^{yz}\ket{\Qzeigenone} &\propto \ket{\Qzeigenthree}, &
Q^{yz}\ket{\Qzeigentwo} &\propto \ket{\Qzeigenthree}.
\end{align}
Hence \(V_x\) and \(V_y\) take the system out of the ground-state manifold of \(H^{(0)}\), so \(P^\dagger V P=0\).
The same reasoning extends to all odd orders: after an odd number of perturbing operators, at least one site on a \(z\)-bond remains excited out of the ground-state manifold of \(H^{(0)}\), so the projected contribution vanishes.

\begin{figure}[tbp]
    \centering
    \includegraphics[width=\columnwidth]{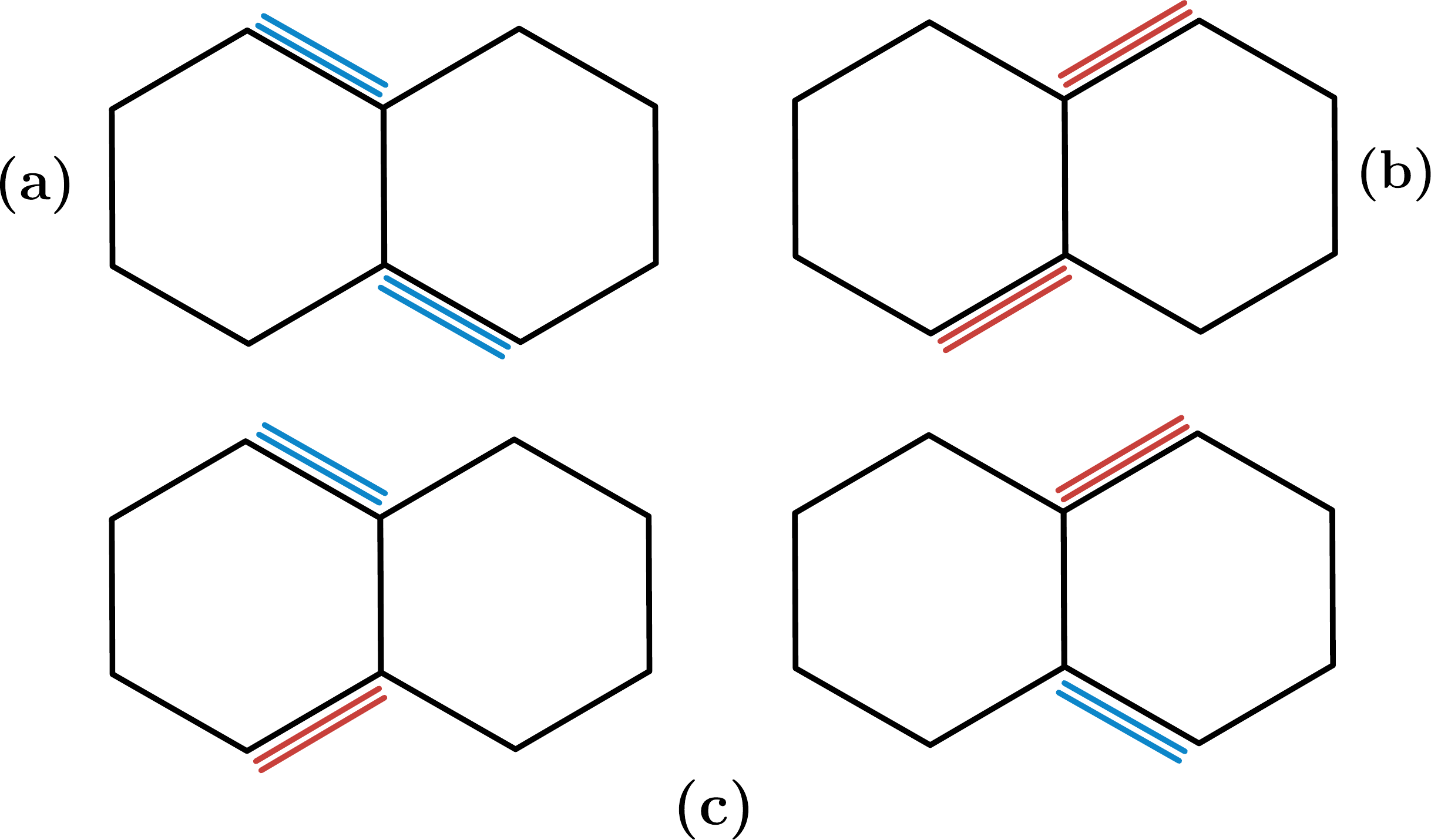}
    \caption{Graphical representation of fourth order perturbative contribution with (a) $\mathcal{O}(J_x^4)$, (b) $\mathcal{O}(J_y^4)$ and (c) $\mathcal{O}(J_x^2J_y^2)$ terms. Both of these terms flip the pseudo-spin state at the $z$ bond which is connecting the two bonds on which the perturbation is applied.}
\label{fig:4th order combined}    
\end{figure}

Turning to even-order contributions, we note that the 2nd-order contribution yields a trivial constant shift, $H^{(2)} \sim \bvec{1}$, and therefore does not lift the degeneracy within the ground-state manifold. The first nontrivial term arises at fourth order,  
\begin{equation}
	H^{(4)} = P^{\dagger} V \left(\frac{1}{E^0 - H^0} R V\right)^3P,
\end{equation}
where \(R=\mathds{1}-P\) projects onto the excited-state subspace of \(H^{(0)}\). At this order, three classes of virtual processes contribute: (i) four successive insertions of \(V_x\), (ii) four successive insertions of \(V_y\), and (iii) two insertions each of \(V_x\) and \(V_y\), as illustrated in Fig.~\ref{fig:4th order combined}.

\section{Majorana mean-field theory} \label{app:partonmft}

\subsection{Mean-field decoupling and Hamiltonian}

Upon performing the mean-field decoupling of the interacting parton Hamiltonian in \eqref{eq:Hamiltonianparton} as schematically indicated in Eq.~\eqref{eq:int-decoupling}, the resulting mean-field Hamiltonian takes the form of Eq.~\eqref{eq:h-mf-pieces}, where $H_{xyz} = - M_A M_B (H_x + H_y + H_z)$ contains the dispersive Majorana fermions of flavors $\alpha = x,y,z$, together with the self-consistency conditions. We have used translational invariance to write $\braket{\Gamma^0_{i,s}} \equiv M_s = \pm 1$.
Explicitly, the contribution $H_x$ associated with the $x$-bond takes the form
\begin{multline}
    \label{eq:meanFieldxbond}
    H_x =
    \iu \frac{J_x}{4}
    \sum_{\langle i,j\rangle\in x}
    \sum_{a,b,c,d}
    D_{abcd}
    \Bigl[
    w^{zz}_{ab}(x)\,\gamma^y_{c,i}\gamma^y_{d,j} \\
    + w^{yz}_{ab}(x)\,\gamma^z_{c,i}\gamma^y_{d,j}
    + w^{zy}_{ab}(x)\,\gamma^y_{c,i}\gamma^z_{d,j}
    + w^{yy}_{ab}(x)\,\gamma^z_{c,i}\gamma^z_{d,j}
    \Bigr],
\end{multline}
where \(D_{abcd}=\pm 1\) if \(c\neq a\), \(d\neq b\) and \(D_{abcd}=0\) otherwise. The sign at non-zero $D_{abcd}$ is \(+ (-)\) if \(a=b,c=d\) \((a\neq b,c\neq d)\). The Hamiltonians \(H_y\) and \(H_z\), describing the corresponding dispersing Majorana modes on \(y\)- and \(z\)-bonds, are obtained from Eq.~\eqref{eq:meanFieldxbond} by the appropriate cyclic relabeling of bond and flavor indices.

Since the bilinear \(\iu \gamma^0_{1,i}\gamma^0_{2,i}=\Gamma_i^0\) is a conserved \(\mathbb Z_2\) quantity taking values \(\pm 1\), the \(\gamma^0\) sector does not describe propagating degrees of freedom at the mean-field level. Instead, it contributes only a static onsite term whose coefficient is determined self-consistently by double contractions of the dispersing Majorana bilinears. Using translation invariance, this contribution can be written as
\begin{equation}
   H_{0}
   =
   \sum_i
   \Bigl(
   M_{B}\,\iu \gamma^0_{1,(i,A)} \gamma^0_{2,(i,A)}
   +
   M_{A}\,\iu \gamma^0_{1,(i,B)} \gamma^0_{2,(i,B)}
   \Bigr)
   \Upsilon_{\mathrm{DC}},
\end{equation}
where the double-contraction coefficient \(\Upsilon_{\mathrm{DC}}\) is given by
\begin{align}
    \Upsilon_{\mathrm{DC}}
    ={}&
    - \frac{J_z}{4}
    \Bigl(
    w^{yy}_{11}(z)w^{xx}_{22}(z)
    - w^{yx}_{12}(z)w^{xy}_{21}(z)\nonumber \\
    +& ~w^{xy}_{11}(z)w^{yx}_{22}(z)
    - w^{xx}_{12}(z)w^{yy}_{21}(z)
    + \cdots
    \Bigr)
    \nonumber\\
    +&~
     \text{similar contributions from the \(x\)- and \(y\)-bonds}.
\end{align}

\subsection{Bilinear forms of the Majorana Hilbert-space constraints}

The local constraint defining the physical parton Hilbert space in Eq.~\eqref{eq:1st constraint} is quartic in the Majorana operators. In order to implement it within a noninteracting-fermion mean-field treatment, it is convenient to impose instead the bilinear conditions
\begin{equation}
    \iu \gamma_a^0 \gamma_a^z + \iu \gamma_a^x \gamma_a^y = 0,
    \qquad a=1,2.
\end{equation}
Indeed, squaring either side gives
\begin{equation}
    \left(\iu \gamma_a^0 \gamma_a^z + \iu \gamma_a^x \gamma_a^y\right)^2
    = 2 - 2\,\gamma_a^0\gamma_a^x\gamma_a^y\gamma_a^z =0,
\end{equation}
so that the bilinear constraint implies the original quartic constraint in Eq.~\eqref{eq:1st constraint}. We note that there are two other equivalent choices for the ``square root'' of Eq.~\eqref{eq:1st constraint} \footnote{corresponding to the three possible pairings of the four Majoranas.}, but we have explicitly verified that they are automatically satisfied whenever the above bilinear condition holds. See also Ref.~\onlinecite{seifertFractionalizedFermiLiquids2018} for a related discussion in the context of Kitaev's four-Majorana fermion-parity constraint.

The second constraint in Eq.~\eqref{eq:2nd constraint} is likewise quartic.
To rewrite this in a bilinear form, it is convenient to use the operators \(\Gamma^\mu=\iu\gamma_1^\mu\gamma_2^\mu\), which leads to
\begin{equation}
\Gamma^0\Gamma^\mu
=
-\gamma_1^0\gamma_2^0\gamma_1^\mu\gamma_2^\mu,
\qquad \mu=x,y,z.
\end{equation}
Hence,
\begin{equation}
\sum_{\mu=x,y,z}\gamma_1^0\gamma_2^0\gamma_1^\mu\gamma_2^\mu=1
\quad\Longleftrightarrow\quad
-\sum_{\mu=x,y,z}\Gamma^0\Gamma^\mu=1.
\end{equation}
Multiplying by \(\Gamma^0\) and using \((\Gamma^0)^2=1\), this is equivalently rewritten as
\begin{equation} \label{eq:constraint_bilinear_final_form}
\sum_{\mu=0,x,y,z}\Gamma^\mu=0.
\end{equation}
Thus, both local Hilbert-space constraints can be imposed at the mean-field level through bilinear operators.

\subsection{Solution of mean-field self-consistency conditions} 

To solve the mean-field self consistency conditions, we exploit translational invariance to work in momentum space.
The Majorana fermions can be expanded as 
\begin{equation}
   \gamma^\mu_{a,(i,s)} = \sqrt{\frac{2}{N_c}} \sum_{k \in  \mathrm{BZ}/2}  \left(\gamma^\mu_{a,(k,s)} \eu^{\iu k \cdot r_i} + \mathrm{h.c.} \right),
\end{equation}
where $s=A,B$ is a sublattice index, and the momentum-space sum extends over only \emph{half} of the Brillouin zone, since the self-conjugate nature of Majorana fermions $\gamma_i^\dagger = \gamma_i$ implies $\gamma_{ k}^\dagger = \gamma_{-{k}}$.
With this, the Hamiltonian can be written in momentum space as
\begin{equation} \label{eq:h-mf-k}
    H_{\mathrm{MF}} = \left(\sum_{k \in\mathrm{BZ}/2} \psi_{k}^{\dagger} H(k)  \psi_{k} \right) - \mathcal{E},
\end{equation}
where $ \psi_{k}$ is a spinor comprised of the complex fermionic annihilation operators
\begin{equation}
    \psi_{\boldsymbol{k}} = \begin{pmatrix} (\gamma^0_{1,(k,A)}) & (\gamma^0_{1,(k,B)})  \cdots 
      (\gamma^z_{2,(k,A)}) & (\gamma^z_{2,(k,B)})\end{pmatrix}^\top,
\end{equation}
and $H(k)$ is, accordingly, a $16 \times 16$-dim. matrix [see the Supplemental Material \cite{suppl} for an explicit definition of $H(k)$], and we further write the constant term as $\mathcal{E}=  3 M_A M_B \Upsilon_{\mathrm{DC}} N_c$, where $N_c = N/2$ is the number of unit cell on the honeycomb lattice (with $N$ sites).

We determine the self-consistent values of the mean-field parameters $w^{\alpha\beta}_{a b}(\delta) = \braket{ \iu \gamma^\alpha_{a,(i,A)} \gamma^\beta_{b,(j,B)}}$ and $M_A,M_B$ along with the average constraints
\begin{multline} \label{eq:methods-constraints}
     \braket{ \iu \gamma_{a,(i,s)}^0 \gamma_{a,(i,s)}^z + \iu \gamma_{a,(i,s)}^x\gamma_{a,(i,s)}^y }_{\mathrm{MF}}=0 \\
    \text{and} \quad \braket{ \sum_{\alpha = {0,x,y,z}} \Gamma^{\alpha}_{(i,s)}}_{\mathrm{MF}}  = 0
\end{multline}
for $s=A,B$ sublattices.
We solve the mean-field equations using an iterative procedure:

\emph{First}, the self-consistency algorithm is initialized with random values for the mean-field parameters. These determine a fermionic mean-field Hamiltonian as given \eqref{eq:h-mf-k}.

\emph{Second}, for each iteration step, the eigenmodes of $H_\mathrm{MF}$ are found by diagonalization of $H(k)$ via some unitary transformation $U$, $U^\dagger_k H(k) U_k = \diag(\varepsilon^{(1)}_k,\dots,\varepsilon^{(16)}_k)$.
The resulting eigenmodes are used to update the mean-field bilinears
\begin{align}
&\langle \iu \gamma^\mu_{a,(i,s)} \gamma^\nu_{b,(i+\delta,s')} \rangle =  \frac{1}{N_c} \sum_{k>0} \left(\iu \langle (\gamma^{\mu}_{a,(k,s)})^{\dagger} \gamma^{\nu}_{b,(k,s')} \rangle \eu^{\iu k \cdot \delta}
 +\mathrm{h.c.}\right) \nonumber \\ 
&= \frac{1}{N_c} \sum_{k>0} \sum_{\rho,\xi=1}^{16} \left( \iu U_{(\nu,b,s'),\rho}\langle \chi_{\xi,k}^{\dagger} \chi_{\rho,k} \rangle  (U^\dagger)_{\xi,(\mu,a,s)} \eu^{\iu k \cdot \delta}+
\mathrm{h.c.}\right), \label{eq:gg-k}
\end{align}
where the $\chi_{\xi,k}$ are annihilation operators for the fermionic eigenmodes of $H_\mathrm{MF}$. At finite temperature \(T=\beta^{-1}\), the eigenmode occupations are taken to be
$\langle \chi_{\xi,k}^{\dagger} \chi_{\rho,k'} \rangle = \delta_{\xi,\rho} \delta_{k,k'} f_{\mathrm{D}}(\beta,\varepsilon^{\xi}_k)$ with $f_{\mathrm{D}}(\beta,\varepsilon) \equiv 1/(\eu^{\beta \varepsilon} +1)$. The zero-temperature mean-field solution is obtained by taking the limit \(\beta\to\infty\).

\emph{Third}, the thus-obtained expectation values provide an updated ansatz for the mean-field parameters, which are updated according to $ w^{(n+1)} = \eta_{\mathrm{mix}}w^{(n)} + \left(1-\eta_{\mathrm{mix}}\right) \langle \iu \gamma^{\mu} \gamma^{\nu} \rangle_{\textrm{calc}}$, where $\eta_{\mathrm{mix}}$ is a heuristically chosen ``mixing parameter'' to improve convergence. Simultaneously, the Lagrange multipliers are modified in a discretized gradient-descent manner $\lambda_a^{(n+1)} = \lambda_a^{(n)} - \eta_{\lambda} (\langle V_a\rangle - \bar{V}_a)$$
$ where $\bar{V}_a \equiv 0$ are the targeted values of the constraint operators appearing in Eq.~\eqref{eq:methods-constraints}, and $\eta_\lambda > 0$ is again some heuristically chosen parameter to optimize convergence.

We iterate the second and third step of the iterative procedure until convergence.
If multiple solutions arise out of the self-consistent procedure, we compare the free energy, $F=E-TS,$ of each of these solutions and select the solution with the lowest free energy.
Here, the internal energy can be written in terms of the eigenmode energies, 
$E = \sum_{\xi=1}^{16} \sum_{k\in \mathrm{BZ}/2} \epsilon_\xi(k) f_\mathrm{D}(\epsilon_i(k)) + \mathcal{E},$ and similarly we use that the entropy of free fermions is expressed as $ S = -\sum_{\xi=1}^{16} \sum_{k\in \mathrm{BZ}/2} \bigl(f_\mathrm{D}(\epsilon_i(k)) \log\left[f_\mathrm{D}(\epsilon_\xi(k))\right] +  \left(1-f_\mathrm{D}(\epsilon_\xi(k))\right) \log\left[1-f_\mathrm{D}(\epsilon_\xi(k))\right] \bigr).$
We have implemented the above procedure numerically, evaluating the momentum-space sums on one half of a discetized Brillouin zone with $40\times 40$ $k$-points, corresponding to a honeycomb lattice of $N_c= 40 \times 40$ unit cells, or $N=2 \times 40^2$ sites.
While we are ultimately interested in the zero-temperature $T=0$ ground state, we work at a small but finite temperature $T=0.01$ to improve convergence and alleviate finite size effects. We have verified that our results are robust upon varying system size.


\end{document}


\title{Supplementary material to ``Exchange-frustrated quadrupoles on the honeycomb lattice: Flavor-wave spectra, classical degeneracies and parton constructions''}

\author{Partha Sarker}
\affiliation{Institute for Theoretical Physics, University of Cologne, 50937 Cologne, Germany}
\author{Han Ma}
\affiliation{CPHT, CNRS, Ecole Polytechnique, Institut Polytechnique de Paris, Palaiseau, France}
\affiliation{Department of Physics and Astronomy, Stony Brook University, \\ Stony Brook, New York 11974, USA}
\affiliation{Perimeter Institute for Theoretical Physics, Waterloo ON N2L 2Y5, Canada}
\author{Urban F.~P.~Seifert}
\affiliation{Institute for Theoretical Physics, University of Cologne, 50937 Cologne, Germany}

\begin{abstract}
In this Supplementary material, we provide further details on the linear-flavor wave spectra as well as additional technical information on our self-consistent Majorana mean-field theory calculations.
\end{abstract}

\date{\today}

\maketitle

\section{Linear flavor-wave theory spectra} \label{sm:lfwt}

\begin{figure}[h!]
    \centering
    \includegraphics[width=0.7\textwidth]{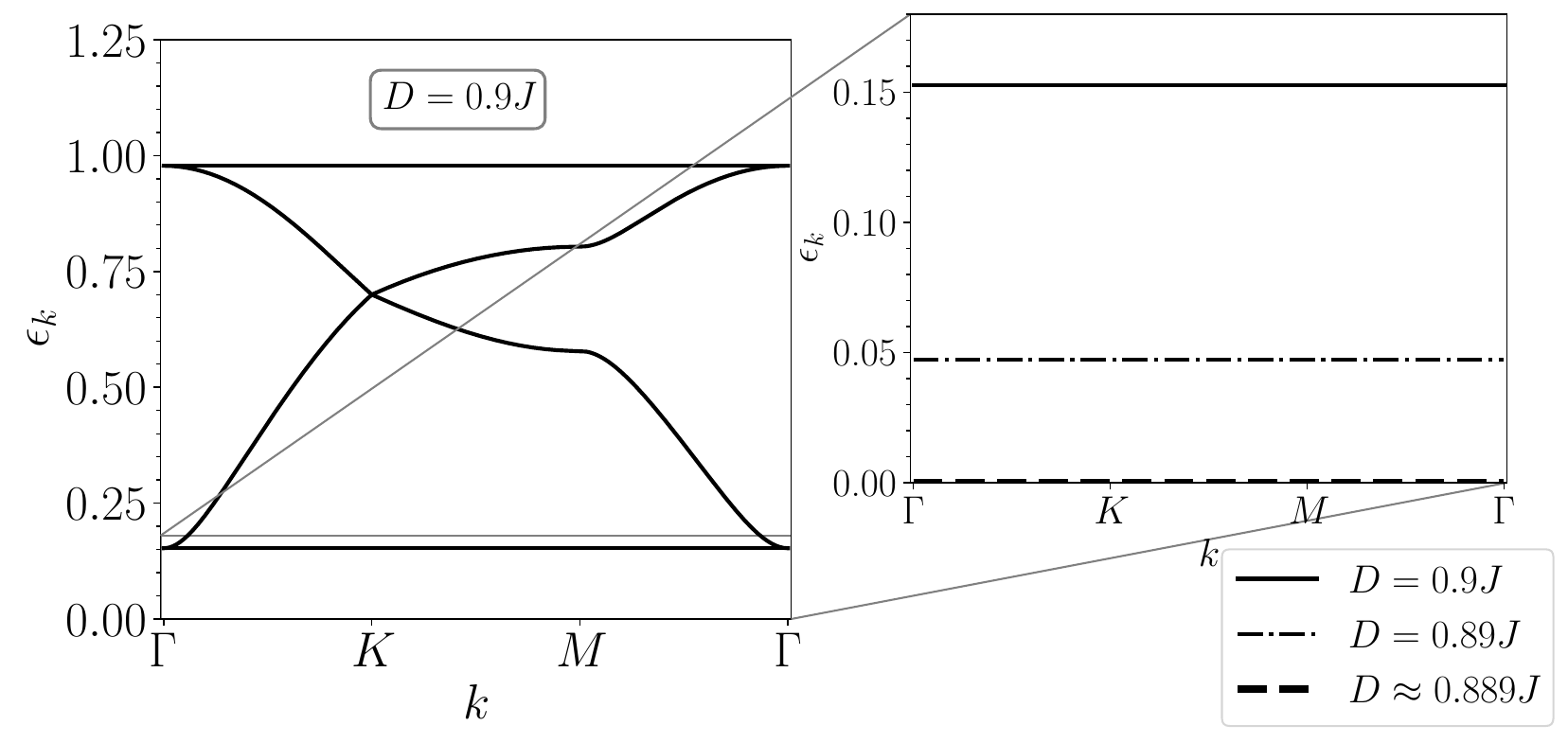}
    \caption{For $[1,1,1]$ SIA deformation as the deformation parameter $D$ is lowered the bandgap closes
    at $D \approx 0.88J$ which signifies the instability of the trivial quadrupolar ordered reference 
    state in presence of strong frustrated quadrupolar Hamiltonian $H_Q$.}
    \label{fig:111anisotropy spectrum}
\end{figure}

\begin{figure}[tbp]
    \centering
    \includegraphics[width=\textwidth]{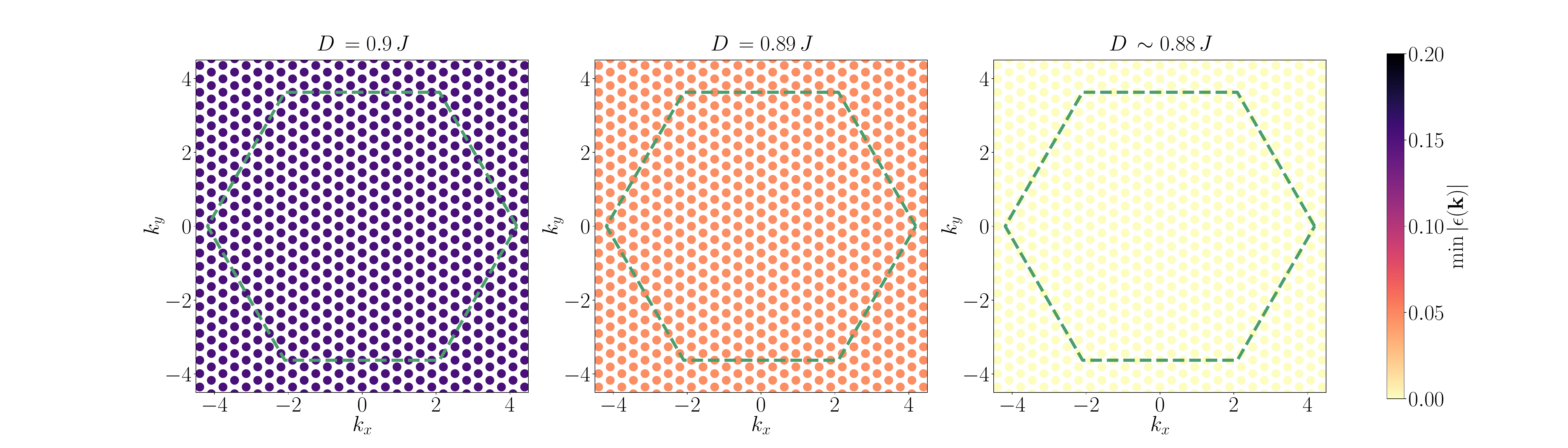}
    \caption{Energy of lowest quasiparticle for the quadrupolar Kitaev model deformed by a  $[111]$ single-ion anisotropy, across the whole $1$ Brillouin zone. The lowest band become 
    gapless at certain value of $D\sim 0.88J$.}
    \label{fig:lowestbandplot111anisotropy}
\end{figure}

\begin{figure}[tbp]
    \centering
    \includegraphics[width=\textwidth]{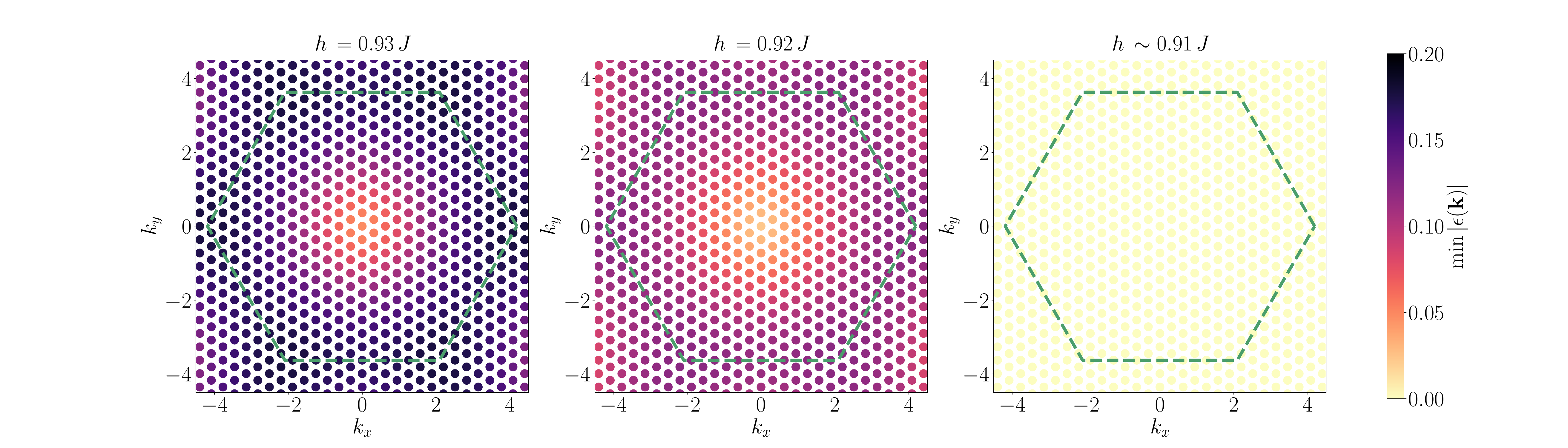}
    \caption{Lowest quasiparticle band in the spin-polarized state, stabilized by a magnetic Zeeamn field along the $[111]$ direction, indicated by the dashed 
    blue lines. The lowest band becomes 
    gapless at certain value of $h\sim 0.91 J$}
    \label{fig:lowestbandplot111field}
\end{figure}

Here, we provide further details on the spectra obtained from linear flavor-wave theory. In Fig.~\ref{fig:111anisotropy spectrum}, we show the lowest flavor-wave band in the trivial ground state stabilized by a single-ion anisotropy along the [111] direction for select values of $D/J$ in vicinity of the critical $D_c \approx 0.88 J$. 
In Fig.~\ref{fig:lowestbandplot111anisotropy}, we show the energy of the lowest quasiparticle band across the hexagonal Brillouin zone near, approaching $D_c$ from above. In Fig.~\ref{fig:lowestbandplot111field}, we similarly show the lowest quasiparticle band as a function of the [111] Zeeman magnetic field (compare also Fig.~2 of the main text).

\section{Additional details on Majorana mean field theory}
\label{sec:AppPartonMFT}
\subsection{Matrix expressions in the \texorpdfstring{$H_{\mathrm{MF}}$}{}}
The contribution to the mean field Hamiltonian $H_{\mathrm{MF}}$ from the quadrupolar Hamiltonian $H_Q$, denoted by $H_{Q,\mathrm{MF}}$, is given by the following block diagonal form
\begin{equation}
    H_{Q,\mathrm{MF}} = \begin{pmatrix}
        \Xi_{00} & \bvec{0}_{4\times 12} \\
        \bvec{0}_{12\times 4} & \Xi_{xyz}
    \end{pmatrix}.
\end{equation}where the $\Xi_{00}$ block coupling $\gamma^0$ Majorana fermions is given by 
\begin{align}
    &\Xi_{00} \nonumber\\
    &= \begin{pmatrix}
        0 & 0 & \iu M_B \Upsilon_{DC} & 0 \\
        0 & 0 & 0 & \iu M_A \Upsilon_{DC} \\
        -\iu M_B \Upsilon_{DC} &  0 & 0 & 0  \\
        0 & -\iu M_A \Upsilon_{DC} &  0 & 0 
    \end{pmatrix},  
\end{align}
and the $12 \times 12$ block $\Xi_{xyz}$ that couples the $\gamma^x,\gamma^y,\gamma^z$ Majorana fermions
\begin{equation}
\Xi_{xyz} 
= {-M_A M_B}\begin{pmatrix}
    \Upsilon_{\w{y}{y}{z}}+\Upsilon_{\w{z}{z}{y}} & \Upsilon_{\w{y}{x}{z}} & \Upsilon_{\w{z}{x}{y}} \\
    \Upsilon_{\w{x}{y}{z}} & \Upsilon_{\w{x}{x}{z}}+\Upsilon_{\w{z}{z}{x}} &\Upsilon_{\w{z}{y}{x}} \\
    \Upsilon_{\w{x}{z}{y}} &  \Upsilon_{\w{y}{z}{x}} & \Upsilon_{\w{x}{x}{y} }+\Upsilon_{\w{y}{y}{x}}
\end{pmatrix}.
\end{equation}
The expressions for the different $4\times4$ blocks such as $\Upsilon_{\w{y}{y}{x}}$ 
\begin{equation}
\Upsilon_{\w{y}{y}{z}}=
 J_z\begin{pmatrix}
        0 & \iu \ws{y}{y}{22}{z} \, e^{\iu k \cdot \delta_z} & 0 & -\iu \ws{y}{y}{21}{z} \, e^{\iu k \cdot \delta_z}\\
        -\iu \ws{y}{y}{22}{z} \, e^{-\iu k \cdot \delta_z} & 0 & \iu \ws{y}{y}{12}{z} \, e^{-\iu k \cdot \delta_z} & 0\\
        0 & -\iu \ws{y}{y}{12}{z} \, e^{\iu k \cdot \delta_z} & 0 & \iu \ws{y}{y}{11}{z} \, e^{\iu k \cdot \delta_z}\\
        \iu \ws{y}{y}{21}{z}\, e^{-\iu k \cdot \delta_z} & 0 & -\iu \ws{y}{y}{11}{z} \, e^{-\iu k \cdot \delta_z} & 0
\end{pmatrix}.   
\end{equation}
where $\delta_z$ is connecting the $A$ sub-lattice site to $B$ sub-lattice site along the $z$ bond. All the other $4 \times 4$ blocks are similar in structure.


\subsection{Effective hopping model for the \texorpdfstring{$\gamma^x,\gamma^y,\gamma^z$}{} Majorana fermions} \label{app:effective hopping model}
For the $\mathcal{M}_z$ broken self-consistent mean field solution, the $\gamma^x$, $\gamma^y$ and $\gamma^z$ fermions disperse along chains of $y$-$z$ bonds, $z$-$x$ bonds and 
$x$-$y$ bonds, respectively, with a staggered amplitude on successive bonds.
Additionally, due to the presence of onsite constraint terms such as $\Gamma^x$ (due to finite Lagrange multipliert $\lambda_{3,s}$), in momentum space we have an effective $4 \times 4$ Hamiltonian for the $\gamma_x$ fermions
\begin{equation}
    H_{\mathrm{eff},x}  = 2 \sum_{k \in \mathrm{BZ}/2 } \psi_{x,k}^{\dagger}   
    \begin{pmatrix}
    0 & T(k) & i\alpha & 0 \\
    T^*(k) & 0 & 0 & -i\alpha \\
    -i\alpha & 0 & 0 & T(k) \\
    0 & i\alpha & T^*(k) & 0
    \end{pmatrix} \psi_{x,k} , 
\end{equation}
where $\psi_{x,k} = \begin{pmatrix}
    \gamma^x_{1,(k,A)} & \gamma^x_{1,(k,B)} & \gamma^x_{2,(k,A)} & \gamma^x_{2,(k,B)}
\end{pmatrix}^\top$ and, at the self-consistent mean-field saddle point, $T = -\iu \, 0.6135 \times M_A M_B (\exp(\iu k \cdot \delta_z) - \exp(\iu k \cdot \delta_y)) /4$ and $\alpha = \lambda_{3,A} = -\lambda_{3,B} \approx 0.05061$.
By diagonalizing this $4 \times 4$ matrix, we evaluate the lowest eigenvalue to be $ 2\sqrt{\alpha^2 + \left|T\right|^2 }$. The minimum value of this eigenvalue i.e., the gap is $\Delta \sim 0.1$. This minimum is achieved along
the $k$ contours $k_x/k_y = \sqrt{3}$ (along $\Gamma$-$M$) in the Brillouin zone. Both the value of the finite gap and the contour in momentum space of smallest gap are consistent with the results of 
Fig.~8 in the main text.